\documentclass[aps,prd,showpacs,onecolumn,superscriptaddress,twoside]{revtex4-2}
\usepackage{amssymb}
\usepackage{mathrsfs}
\usepackage{amsfonts}
\usepackage{epstopdf}
\usepackage{bm,color,bbm}
\usepackage{amsmath}
\usepackage{graphicx}
\usepackage{natbib}
\usepackage[hidelinks]{hyperref}
\usepackage{caption}
\usepackage{subcaption}

\usepackage{fancyhdr}
\providecommand{\U}[1]{\protect\rule{.1in}{.1in}}

\begin{document}
\title{\textbf{Efficiency, Curvature, and Complexity of Quantum Evolutions for Qubits
in Nonstationary Magnetic Fields}}
\author{\textbf{Carlo Cafaro}$^{1}$ and \textbf{James Schneeloch}$^{2}$}
\affiliation{$^{1}$University at Albany-SUNY, Albany, NY 12222, USA}
\affiliation{$^{2}$Air Force Research Laboratory, Information Directorate, Rome, New York,
13441, USA}

\begin{abstract}
In optimal quantum-mechanical evolutions, motion can take place along paths of
minimal length within an optimal time frame. Alternatively, optimal evolutions
may occur along established paths without any waste of energy resources and
achieving $100\%$ speed efficiency. Unfortunately, realistic physical
scenarios often lead to less-than-ideal evolutions that demonstrate suboptimal
efficiency, nonzero curvature, and a high level of complexity.

In this paper, we provide an exact analytical expression for the curvature of
a quantum evolution pertaining to a two-level quantum system subjected to
various time-dependent magnetic fields. Specifically, we examine the dynamics
produced by a two-parameter nonstationary Hermitian Hamiltonian with unit
speed efficiency. To enhance our understanding of the physical implications of
the curvature coefficient, we analyze the curvature behavior in relation to
geodesic efficiency, speed efficiency, and the complexity of the quantum
evolution (as described by the ratio of the difference between accessible and
accessed Bloch-sphere volumes for the evolution from initial to final state to
the accessible volume for the given quantum evolution). Our findings indicate
that, generally, efficient quantum evolutions exhibit lower complexity
compared to inefficient ones. However, we also note that complexity transcends
mere length. In fact, longer paths that are sufficiently curved can
demonstrate a complexity that is less than that of shorter paths with a lower
curvature coefficient.

\end{abstract}

\maketitle

\thispagestyle{fancy}

\section{Introduction}

It is recognized that geometric reasoning can provide significant insights in
the realm of quantum physics \cite{feynman57}. For example, when concentrating
on pure states, we can pinpoint Hamiltonian operators that facilitate quantum
evolutions at the maximum achievable rate
\cite{bender07,bender09,ali09,cafarocqg,hete23,cc23,cc24}. These quantum
dynamical trajectories, referred to as Hamiltonian curves, adhere to geodesic
paths on the metricized manifolds of quantum states, as the states progress
according to defined physical evolutions. A key quantity employed to measure
the deviation from the ideal geodesic evolution on a manifold of pure states
is the so-called geodesic efficiency $\eta_{\mathrm{GE}}$, which was
introduced by Anandan and Aharonov in Ref. \cite{anandan90}. This quantity
serves as a global measure since it is assessed over a finite time interval.
It is articulated in terms of the ratio between two lengths, $s_{0}$ and $s$.
The length $s_{0}$ represents the geodesic distance, which is the length of
the shortest path connecting the initial and final states ($\left\vert
A\right\rangle $ and $\left\vert B\right\rangle $, respectively) as measured
with respect to the Fubini-Study metric (where $s_{0}=2s_{\mathrm{FS}}$, with
$s_{\mathrm{FS}}$ denoting the Fubini-Study distance). Conversely, the length
$s$ characterizes the actual path traversed by the quantum system.
Specifically, $s$ is influenced by the energy uncertainty of the system, which
is governed by either a stationary or a time-varying Hamiltonian.

\medskip

In geometric quantum mechanics, the analysis of parameter estimation
introduces the concept of curvature of a quantum Schr\"{o}dinger trajectory,
as presented in Ref. \cite{brody96}. This concept serves as a generalization
of the curvature notion found in classical exponential families of
distributions utilized in statistical mechanics. In Ref. \cite{brody96}, the
curvature of a curve is defined through a suitably introduced squared
acceleration vector associated with the curve. Moreover, this curvature
quantifies the parametric sensitivity relevant to the parametric estimation
problem under investigation \cite{brody13}. In Ref. \cite{laba17}, Laba and
Tkachuk established definitions for curvature and torsion coefficients
pertaining to quantum evolutions of pure states that undergo time-independent
Hamiltonian evolution. In particular, by concentrating on single-qubit quantum
states, they demonstrated that curvature serves as a measure of how much the
dynamically evolving state vector deviates from the geodesic line on the Bloch
sphere. Utilizing the Frenet-Serret apparatus concept and partially drawing
from the research conducted by Laba and Tkachuk as referenced in
\cite{laba17}, an alternative geometric framework to define the bending and
twisting of quantum curves represented by dynamically evolving state vectors
within a quantum context was introduced in Refs. \cite{alsing24A,
alsing24B,alsing24C}. In particular, it was proposed a quantum adaptation of
the Frenet-Serret apparatus for a quantum trajectory in projective Hilbert
space, which is delineated by a parallel-transported pure quantum state that
evolves unitarily under a stationary (or, alternatively, nonstationary)
Hamiltonian that governs the Schr\"{o}dinger equation. In the case of
nonstationary Hamiltonian evolutions, the authors reported in Ref.
\cite{alsing24B} that the time-varying framework revealed a more intricate
structure from a statistical perspective compared to the stationary framework
\cite{alsing24A}. To effectively demonstrate the relevance of their construct,
the authors applied it to an exactly solvable time-dependent two-state Rabi
problem characterized by a sinusoidal oscillating time-dependent potential.
Within this framework, they illustrated that the analytical formulations for
the curvature and torsion coefficients can, in principle, be entirely
described using only two real three-dimensional vectors: the Bloch vector that
defines the quantum system and the externally applied time-varying magnetic
field. Although they established that the torsion is identically zero for any
arbitrary time-dependent single-qubit Hamiltonian evolution, they only
examined the temporal behavior of the curvature coefficient numerically across
various dynamical scenarios, including both off-resonance and on-resonance
conditions, as well as strong and weak driving configurations. In Ref.
\cite{cafaropra25}, the first exact analytical analysis of the curvature of
the trajectory of a two-level quantum system subjected to a time-varying
magnetic field was recorded in the literature. Nonetheless, only one temporal
profile of the magnetic field was examined.

\medskip

The importance of the concepts of length and volume is vital in establishing
complexity within a quantum mechanical context. In the field of quantum
physics, there are numerous interpretations of complexity. For instance, one
might refer to the complexity linked to a quantum state (state complexity,
\cite{chapman18,iaconis21}), the complexity present in a quantum circuit
(circuit complexity, \cite{nielsen,fernando21}), or the complexity associated
with an operator (operator complexity, \cite{parker19,roy23,liu23,caputa22}).
Although these measures of complexity possess unique attributes, they are
united by a fundamental principle: the complexity of a composite system
generally escalates in relation to the number of essential components needed
for its assembly \cite{lof77,ay08,vijay22}. Depending on the situation, this
correlation can frequently be expressed through geometrically intuitive
concepts such as lengths and volumes. In the area of theoretical computer
science, Kolmogorov asserted in Ref. \cite{kolmogorov68} that the complexity
of a sequence can be measured by the length of the shortest Turing machine
program that can produce it. Furthermore, in the sphere of information theory,
Rissanen proposed in Refs. \cite{rissanen78,rissanen86} that the average
minimal code length of a collection of messages acts as an indicator of the
complexity of that collection.

Returning to the geometric quantifiers of complexity, state complexity
pertains to the intricacy of a quantum state, defined in terms of the minimal
local unitary circuit that can generate the state from a fundamental (i.e.,
factorizable) reference quantum state. Notably, a geometric perspective on
state complexity was first introduced during the investigation of quantum
state complexity in continuous many-body systems. Specifically, the state
complexity of a target state, obtained by applying a sequence of parametrized
unitary operators to a source state, was depicted as the length of the
shortest path, as determined by the Fubini-Study metric, that corresponds to a
valid realization of the unitary operator \cite{chapman18}. Quantum circuits
are composed of quantum gates that act on quantum states. In particular,
circuit complexity refers to the minimum number of primitive gates necessary
to create a circuit that executes a transformation on a designated quantum
state \cite{nielsen,fernando21}. This complexity is fundamentally discrete and
serves as an important metric for researchers involved in the practical
development of quantum circuits from basic gates. Importantly, the notions of
state complexity and circuit complexity are interrelated; state complexity
relates to the least complex unitary operator that connects the source and
target states. The geometric characterization of circuit complexity was
proposed by Nielsen and his associates in Refs. \cite{mike06,mike06B,mike08}.
Within this geometric framework, the circuit complexity linked to a unitary
operator $U$ is inherently continuous and corresponds to the length of the
shortest geodesic that connects the identity operator to $U$ within the
unitary group. The lengths of these geodesic paths act as a lower limit for
the minimum number of quantum gates necessary to create the unitary operator
$U$. Additionally, operator complexity relates to the temporal growth in the
size of an operator as it progresses under the Heisenberg or Lindblad dynamics
for closed and open quantum systems, respectively. When evaluated in the
context of the Krylov basis, this operator complexity is termed Krylov
complexity \cite{parker19}. For completeness, we remark that the size of an
evolving operator is measured by the average size of the basis operators in
its expansion. Basis operators are simple local operators such as single-site
Pauli operators on a spin-chain and, in addition, their size is specified by
the number of sites on which they act nontrivially. In Ref. \cite{caputa22},
the time evolution of the displacement operator during the unitary evolution
of many-body quantum systems characterized by symmetries demonstrated that the
expectation value of the Krylov complexity operator is equivalent to the
volume of the corresponding classical phase space. The connection between
Krylov complexity and volume was further investigated within particular
quantum field theoretic frameworks. Importantly, the scaling of this
complexity with volume was validated in Ref. \cite{roy23} by illustrating that
Krylov complexity is associated with the average particle number. This
connection was effectively justified by utilizing the proportionality between
volume and average particle number.

In Refs. \cite{npb25,epjplus25}, an alternative measure of complexity was
introduced for examining the complexities of both time-optimal and time
sub-optimal quantum Hamiltonian evolutions that connect arbitrary source and
target states on the Bloch sphere, utilizing the Fubini-Study metric. The
notion of complexity was articulated in terms of the accessed (i.e., partial)
and accessible (i.e., total) parametric volumes of the areas on the Bloch
sphere that delineate the quantum mechanical evolution from the source to the
target states. Specifically, by concentrating on time optimal and time
suboptimal evolutions, the characteristics of complexity were juxtaposed with
those of path length, geodesic efficiency, and the curvature coefficient that
define the evolutions. Generally, it was determined that efficient quantum
evolutions exhibit lower complexity compared to inefficient ones.
Nevertheless, the authors also noted in Refs.\cite{npb25,epjplus25} that
complexity transcends mere length. In fact, longer paths that are sufficiently
curved can demonstrate a complexity that is less than that of shorter paths
with a lower curvature coefficient. However, the analysis was confined to
stationary Hamiltonian evolutions.

\medskip

Driven by the lack of research on the geometric characterization of
nonstationary Hamiltonian evolutions in two-level quantum systems,
particularly regarding efficiency, curvature, and complexity, this paper seeks
to fill this gap by addressing fundamental questions, including:

\begin{enumerate}
\item[{[i]}] The analytical solution of Schr\"{o}dinger's evolution equation
for two-level systems in the presence of time-dependent magnetic field
configurations can be quite complex. Is it possible to identify a category of
time-varying magnetic field configurations for which precise analytical
solutions can be derived?

\item[{[ii]}] Relative phases hold physical significance, exhibiting
observable effects in quantum mechanics. Can we establish a general method to
connect these relative phases to specific magnetic field configurations that
define the nonstationary Hamiltonians of interest?

\item[{[iii]}] Having established the relationship between generally
time-varying phases and magnetic field configurations, can we measure the
impact of their appropriately chosen temporal behavior on the path lengths,
energy dissipation, and degree of trajectory curvature when examining the
quantum evolution between fixed initial and final states on the Bloch sphere?

\item[{[iv]}] With a clear understanding of how relative phases and magnetic
field configurations influence these geometric characteristics of quantum
dynamical trajectories (i.e., geodesic efficiency, speed efficiency, and
curvature coefficient), can we grasp how they affect the complexity of the
dynamical evolutions in terms of the accessed and accessible volumes of
parametric regions of the Bloch sphere?
\end{enumerate}

Addressing these issues is significant for various reasons within the domain
of quantum information and computation. For example, a thorough quantitative
comprehension of these matters can assist in the development of appropriate
time-dependent quantum driving strategies that are capable of transferring a
source state to a target state in the least amount of time, at optimal speed,
with minimal energy resource waste, and ultimately, with reduced complexity.

\medskip

The layout of the rest of this paper is as follows. In Section II, we present
the fundamental components necessary for comprehending the concepts of
geodesic efficiency and speed efficiency \cite{uzdin12} in quantum-mechanical
evolutions. In Section III, we advance to the characterization of the
geometric features of quantum evolutions by introducing the notions of
curvature coefficient and the complexity of dynamical trajectories traced by
state vectors, whose dynamics are dictated by nonstationary Hamiltonians. In
Section IV, we introduce a two-parameter family of time-varying Hamiltonians,
which we aim to analyze geometrically. Specifically, we concentrate on the
time-dependent configurations of the magnetic fields that define the
Hamiltonians, as well as the temporal behavior of the phase that determines
the relative phase factor involved in the decomposition of the evolving state
in terms of the computational basis state vectors. In Section V, we utilize
the concepts introduced in Sections II and III to characterize the quantum
evolutions discussed in Section IV. In particular, by focusing on the temporal
dependence of the aforementioned phase, we examine five distinct scenarios: i)
no growth; ii) linear growth; iii) quadratic growth; iv) exponential growth;
v) exponential decay. Our summary of findings, along with our concluding
remarks, can be found in Section V. Finally, technical details appear in
Appendix A.

\section{Efficiency}

In this section, we outline the essential elements required to understand the
principles of geodesic efficiency and speed efficiency in quantum-mechanical evolutions.

\subsection{Geodesic Efficiency}

Consider the evolution of a state vector $\left\vert \psi\left(  t\right)
\right\rangle $ as described by the time-dependent Schr\"{o}dinger equation,
$i\hslash\partial_{t}\left\vert \psi\left(  t\right)  \right\rangle
=\mathrm{H}\left(  t\right)  \left\vert \psi\left(  t\right)  \right\rangle $,
within the interval $t_{A}\leq t\leq t_{B}$. Consequently, the geodesic
efficiency $\eta_{\mathrm{GE}}$ for this quantum evolution is a scalar
quantity that remains constant over time (global) and is defined within the
range of $0\leq\eta_{\mathrm{GE}}\leq1$. It equals \cite{anandan90,cafaro20}%
\begin{equation}
\eta_{\mathrm{GE}}\overset{\text{def}}{=}\frac{s_{0}}{s}=\frac{2\arccos\left[
\left\vert \left\langle A|B\right\rangle \right\vert \right]  }{2\int_{t_{A}%
}^{t_{B}}\frac{\Delta E\left(  t\right)  }{\hslash}dt}\text{.}
\label{efficiency}%
\end{equation}
The quantity $s_{0}$ represents the distance along the shortest geodesic path
connecting the initial state $\left\vert A\right\rangle \overset{\text{def}%
}{=}$ $\left\vert \psi\left(  t_{A}\right)  \right\rangle $ and the final
state $\left\vert B\right\rangle \overset{\text{def}}{=}\left\vert \psi\left(
t_{B}\right)  \right\rangle $ within the complex projective Hilbert space.
Moreover, the quantity $s$ in Eq. (\ref{efficiency}) signifies the distance
along the dynamical trajectory $\gamma\left(  t\right)  :t\mapsto\left\vert
\psi\left(  t\right)  \right\rangle $ that corresponds to the evolution of the
state\textbf{ }vector $\left\vert \psi\left(  t\right)  \right\rangle $ for
$t_{A}\leq t\leq t_{B}$. Clearly, a geodesic quantum evolution characterized
by $\gamma\left(  t\right)  =\gamma_{\mathrm{geo}}\left(  t\right)  $ is
established through the equation $\eta_{\mathrm{GE}}^{(\gamma_{\mathrm{geo})}%
}=1$. By concentrating on the numerator in Eq. (\ref{efficiency}), we observe
that it specifies the angle between the unit state vectors $\left\vert
A\right\rangle $ and $\left\vert B\right\rangle $, which is equivalent to the
Wootters distance \cite{wootters81}. Specifically, we define $\rho
_{A}\overset{\text{def}}{=}\left\vert A\right\rangle \left\langle A\right\vert
=(\mathbf{1+a}\cdot\mathbf{\boldsymbol{\sigma}})/2$ and $\rho_{B}%
\overset{\text{def}}{=}\left\vert B\right\rangle \left\langle B\right\vert
=(\mathbf{1+b}\cdot\mathbf{\boldsymbol{\sigma}})/2$, where $\mathbf{a}$ and
$\mathbf{b}$ are unit vectors such that $\mathbf{a}\cdot$ $\mathbf{b}%
=\cos(\theta_{AB})$. Consequently, it follows that $s_{0}=\theta_{AB}$, given
that $\left\vert \left\langle A|B\right\rangle \right\vert ^{2}=\mathrm{tr}%
\left(  \rho_{A}\rho_{B}\right)  +2\sqrt{\det(\rho_{A})\det(\rho_{B}%
)}=(1+\mathbf{a}\cdot\mathbf{b})/2=\cos^{2}\left(  \theta_{AB}/2\right)  $.
Evidently, $\mathbf{\boldsymbol{\sigma}}\overset{\text{def}}{=}\left(
\sigma_{x}\text{, }\sigma_{y}\text{, }\sigma_{z}\right)  $ represents the
vector operator defined by the standard Pauli operators. Conversely, the
denominator in Eq. (\ref{efficiency}) denotes the integral of the
infinitesimal distance $ds\overset{\text{def}}{=}2\left[  \Delta E\left(
t\right)  /\hslash\right]  dt$ along the evolution curve in ray space
\cite{anandan90}. The term $\Delta E\left(  t\right)  \overset{\text{def}%
}{=}\left[  \left\langle \psi|\mathrm{H}^{2}\left(  t\right)  |\psi
\right\rangle -\left\langle \psi|\mathrm{H}\left(  t\right)  |\psi
\right\rangle ^{2}\right]  ^{1/2}$ signifies the energy uncertainty of the
system, articulated as the square root of the dispersion of $\mathrm{H}\left(
t\right)  $. Significantly, Anandan and Aharonov demonstrated that the
infinitesimal distance $ds\overset{\text{def}}{=}2\left[  \Delta E\left(
t\right)  /\hslash\right]  dt$ is related to the Fubini-Study infinitesimal
distance $ds_{\text{\textrm{FS}}}$ through the relationship \cite{anandan90},%
\begin{equation}
ds_{\text{\textrm{FS}}}^{2}\left(  \left\vert \psi\left(  t\right)
\right\rangle \text{, }\left\vert \psi\left(  t+dt\right)  \right\rangle
\right)  \overset{\text{def}}{=}4\left[  1-\left\vert \left\langle \psi\left(
t\right)  |\psi\left(  t+dt\right)  \right\rangle \right\vert ^{2}\right]
=4\frac{\Delta E^{2}\left(  t\right)  }{\hslash^{2}}dt^{2}+\mathcal{O}\left(
dt^{3}\right)  \text{,} \label{relation}%
\end{equation}
where $\mathcal{O}\left(  dt^{3}\right)  $ denotes an infinitesimal quantity
of an order that is equal to or greater than $dt^{3}$. From the relationship
between $ds_{\mathrm{FS}}$ and $ds$, it can be concluded that $s$ is
proportional to the time integral of $\Delta E$. Furthermore, $s$ represents
the distance calculated by the Fubini-Study metric throughout the evolution of
the quantum system in ray space. It is important to emphasize that when the
actual dynamical curve corresponds to the shortest geodesic path connecting
$\left\vert A\right\rangle $ and $\left\vert B\right\rangle $, $s$ is equal to
$s_{0}$, and the geodesic efficiency $\eta_{\mathrm{GE}}$ in Eq.
(\ref{efficiency}) equals one. Evidently, $\pi$ represents the minimal
distance between two orthogonal pure states within ray space. Notably, we
observe that by defining \textrm{H}$\left(  t\right)  \overset{\text{def}%
}{=}h_{0}\left(  t\right)  \mathbf{1+h}\left(  t\right)  \cdot
\mathbf{\boldsymbol{\sigma}}$ and $\rho\left(  t\right)  \overset{\text{def}%
}{=}(\mathbf{1+a}\left(  t\right)  \cdot\mathbf{\boldsymbol{\sigma})/}2$ for
$t_{A}\leq t\leq t_{B}$, the energy uncertainty simplifies to $\Delta E\left(
t\right)  =\sqrt{\mathbf{h}^{2}-\left[  \mathbf{a}\left(  t\right)
\cdot\mathbf{h}\right]  ^{2}}$. Ultimately, the geodesic efficiency expressed
in in Eq. (\ref{efficiency}) can be reformulated as%
\begin{equation}
\eta_{\mathrm{GE}}=\frac{2\arccos\left(  \sqrt{\frac{1+\mathbf{a}%
\cdot\mathbf{b}}{2}}\right)  }{\int_{t_{A}}^{t_{B}}\frac{2}{\hslash}%
\sqrt{\mathbf{h}^{2}-\left[  \mathbf{a}\left(  t\right)  \cdot\mathbf{h}%
\right]  ^{2}}dt}\text{,} \label{jap}%
\end{equation}
where $\mathbf{a}\left(  t_{A}\right)  \overset{\text{def}}{=}\mathbf{a}$ and
$\mathbf{a}\left(  t_{B}\right)  =\mathbf{b}$ in Eq. (\ref{jap}).
Intriguingly, assuming $\mathrm{H}\left(  t\right)  \overset{\text{def}%
}{=}\mathbf{h}\left(  t\right)  \cdot\mathbf{\boldsymbol{\sigma}}$ and putting
$\mathbf{h}=\left[  \mathbf{h}\cdot\mathbf{a}\right]  \mathbf{a}+\left[
\mathbf{h}-(\mathbf{h}\cdot\mathbf{a})\mathbf{a}\right]  =\mathbf{h}%
_{\shortparallel}+\mathbf{h}_{\perp}$ (where, of course, $\mathbf{a}%
=\mathbf{a}\left(  t\right)  $ in the decomposition of $\mathbf{h}$), the
geodesic efficiency $\eta_{\mathrm{GE}}$ in Eq. (\ref{jap}) becomes (setting
$\hslash=1$, which results in $\mathbf{h}$ being expressed not in units of
energy, but rather in units of frequency)%
\begin{equation}
\eta_{\mathrm{GE}}=\frac{\arccos\left(  \sqrt{\frac{1+\mathbf{a}%
\cdot\mathbf{b}}{2}}\right)  }{\int_{t_{A}}^{t_{B}}h_{\bot}(t)dt}\text{.}
\label{goodyo1}%
\end{equation}
Consequently, from Eq. (\ref{goodyo1}), it is evident that $\eta_{\mathrm{GE}%
}$ is solely dependent on $h_{\bot}(t)$. Interestingly, by maintaining
$\hslash=1$, we observe that Feynman's geometric evolution equation
\cite{feynman57} $d\mathbf{a}/dt=2\mathbf{h}\times\mathbf{a}$, which pertains
to the time-dependent unit Bloch vector $\mathbf{a}=\mathbf{a}\left(
t\right)  $, can be interpreted as a local expression of the Anandan-Aharonov
relation $s=2\int\Delta E\left(  t\right)  dt$. In fact, from $d\mathbf{a}%
/dt=2\mathbf{h}\times\mathbf{a}$, we derive $da^{2}=d\mathbf{a\cdot
}d\mathbf{a}=4h_{\bot}^{2}dt^{2}$. Likewise, from $s=2\int\Delta E\left(
t\right)  dt$, we can deduce $ds=2h_{\bot}dt$. Thus, by merging these two
differential equations, we ultimately arrive at $da=\sqrt{d\mathbf{a\cdot
}d\mathbf{a}}=2h_{\bot}dt=ds$.

After this concise overview of geodesic efficiency, we are now prepared to
introduce the notion of speed efficiency.

\subsection{Speed Efficiency}

Suitable families of nonstationary Hamiltonians that can produce predetermined
dynamical trajectories with minimal energy resource waste were originally
introduced in Ref. \cite{uzdin12}. While these trajectories are efficient in
terms of energy, they do not typically represent geodesic paths of the
shortest length. The condition for minimal energy waste is achieved when no
energy is expended on segments of the Hamiltonian $\mathrm{H}=\mathrm{H}%
\left(  t\right)  $ that do not effectively guide the system. In other words,
all the energy available, as indicated by the spectral norm of the Hamiltonian
$\left\Vert \mathrm{H}\right\Vert _{\mathrm{SP}}$, is converted into the
system's evolution speed $v_{\mathrm{H}}(t)$ $\overset{\text{def}%
}{=}(2/\hslash)\Delta E\left(  t\right)  $, where $\Delta E\left(  t\right)  $
denotes the energy uncertainty. More explicitly, Uzdin's speed efficiency
$\eta_{\mathrm{SE}}$ refers to a time-dependent (local) scalar quantity that
satisfies the condition $0\leq\eta_{\mathrm{SE}}\leq1$. It is defined as%
\begin{equation}
\eta_{\mathrm{SE}}\left(  t\right)  \overset{\text{def}}{=}\frac
{\Delta\mathrm{H}_{\rho}}{\left\Vert \mathrm{H}\right\Vert _{\mathrm{SP}}%
}=\frac{\sqrt{\mathrm{tr}\left(  \rho\mathrm{H}^{2}\right)  -\left[
\mathrm{tr}\left(  \rho\mathrm{H}\right)  \right]  ^{2}}}{\max\left[
\sqrt{\mathrm{eig}\left(  \mathrm{H}^{\dagger}\mathrm{H}\right)  }\right]
}\text{.} \label{se1}%
\end{equation}
While $\Delta\mathrm{H}_{\rho}=\Delta E\left(  t\right)  $ and $\rho
=\rho\left(  t\right)  $ represents the density operator that characterizes
the quantum system at time $t$, the term $\left\Vert \mathrm{H}\right\Vert
_{\mathrm{SP}}$ found in the denominator of Eq. (\ref{se1}) is defined as
$\left\Vert \mathrm{H}\right\Vert _{\mathrm{SP}}\overset{\text{def}}{=}%
\max\left[  \sqrt{\mathrm{eig}\left(  \mathrm{H}^{\dagger}\mathrm{H}\right)
}\right]  $. This term is known as the spectral norm $\left\Vert
\mathrm{H}\right\Vert _{\mathrm{SP}}$ of the Hamiltonian operator \textrm{H},
serving as a measure of the magnitude of bounded linear operators. It is
calculated as the square root of the maximum eigenvalue of the operator
$\mathrm{H}^{\dagger}\mathrm{H}$, where $\mathrm{H}^{\dagger}$ signifies the
Hermitian conjugate of $\mathrm{H}$. Concentrating on two-level quantum
systems and presuming the nonstationary Hamiltonian in Eq. (\ref{se1}) is
represented by $\mathrm{H}\left(  t\right)  \overset{\text{def}}{=}%
h_{0}\left(  t\right)  \mathbf{1}+\mathbf{h}\left(  t\right)  \cdot
\mathbf{\boldsymbol{\sigma}}$, we find that the speed efficiency
$\eta_{\mathrm{SE}}$ in Eq. (\ref{se1}) can be effectively articulated as%
\begin{equation}
\eta_{\mathrm{SE}}\left(  t\right)  \overset{\text{def}}{=}\frac
{\sqrt{\mathbf{h}^{2}-(\mathbf{a}\cdot\mathbf{h})^{2}}}{\left\vert
h_{0}\right\vert +\sqrt{\mathbf{h}^{2}}}\text{.} \label{se2}%
\end{equation}
While $\mathbf{a=a}\left(  t\right)  $ in Eq. (\ref{se2}) represents the
instantaneous unit Bloch vector characterizing the qubit state of the system,
the set $\mathrm{eig}\left(  \mathrm{H}^{\dagger}\mathrm{H}\right)  $ in the
definition of $\left\Vert \mathrm{H}\right\Vert _{\mathrm{SP}}$ is equivalent
to $\mathrm{eig}\left(  \mathrm{H}^{\dagger}\mathrm{H}\right)  =\left\{
\lambda_{\mathrm{H}^{\dagger}\mathrm{H}}^{\left(  +\right)  }%
\overset{\text{def}}{=}(h_{0}+\sqrt{\mathbf{h}^{2}})^{2}\text{, }%
\lambda_{\mathrm{H}^{\dagger}\mathrm{H}}^{\left(  -\right)  }%
\overset{\text{def}}{=}(h_{0}-\sqrt{\mathbf{h}^{2}})^{2}\right\}  $. It is
important to note that the eigenvalues of $\mathrm{H}\left(  t\right)
\overset{\text{def}}{=}h_{0}\left(  t\right)  \mathbf{1}+\mathbf{h}\left(
t\right)  \cdot\mathbf{\boldsymbol{\sigma}}$ are given by $E_{\pm
}\overset{\text{def}}{=}h_{0}\pm\sqrt{\mathbf{h\cdot h}}$. Consequently, the
term $h_{0}=(E_{+}+E_{-})/2$ signifies the mean value of the two energy
levels. Moreover, $\sqrt{\mathbf{h}^{2}}=(E_{+}-E_{-})/2$ is directly
proportional to the energy difference $E_{+}-E_{-}$ between the two energy
states $E_{\pm}$, where $E_{+}$ is greater than or equal to $E_{-}$. Lastly,
for a time-dependent Hamiltonian $\mathrm{H}\left(  t\right)
\overset{\text{def}}{=}\mathbf{h}\left(  t\right)  \cdot
\mathbf{\boldsymbol{\sigma}}$ that satisfies $\mathbf{a}(t)\cdot
\mathbf{h}\left(  t\right)  =0$ at every moment in time $t$, the speed
efficiency $\eta_{\mathrm{SE}}\left(  t\right)  $ equals one. Consequently,
quantum evolution for a time-dependent Hamiltonian with stationary eigenvalues
occurs without any depletion of energy resources. Notably, for $\mathrm{H}%
\left(  t\right)  \overset{\text{def}}{=}\mathbf{h}\left(  t\right)
\cdot\mathbf{\boldsymbol{\sigma}}$ and by defining $\mathbf{h}=(\mathbf{h}%
\cdot\mathbf{a})\mathbf{a}+\left[  \mathbf{h}-(\mathbf{h}\cdot\mathbf{a}%
)\mathbf{a}\right]  =\mathbf{h}_{\shortparallel}+\mathbf{h}_{\perp}$ with
$\mathbf{h}_{\shortparallel}\cdot\mathbf{h}_{\perp}=0$, $\eta_{\mathrm{SE}%
}\left(  t\right)  $ in Eq. (\ref{se2}) can be entirely represented by means
of the parallel (i.e., $\mathbf{h}_{\shortparallel}\overset{\text{def}%
}{=}h_{\shortparallel}\hat{h}_{\shortparallel}$) and transverse (i.e.,
$\mathbf{h}_{\perp}\overset{\text{def}}{=}h_{\perp}\hat{h}_{\perp}$)
components of the magnetic\ field vector $\mathbf{h}$ as%
\begin{equation}
\eta_{\mathrm{SE}}\left(  t\right)  =\frac{h_{\bot}(t)}{\sqrt{h_{\bot}%
^{2}(t)+h_{\shortparallel}^{2}(t)}}\text{.} \label{eqtick}%
\end{equation}
From Eq. (\ref{eqtick}), it is clear that $\eta_{\mathrm{SE}}\left(  t\right)
=1$ if and only if $h_{\shortparallel}(t)=0$ for all $t.$ In other terms, this
condition holds true if and only if the orthogonality condition $\mathbf{a}%
\left(  t\right)  \cdot\mathbf{h}\left(  t\right)  =0$ is fulfilled for every
$t$.

After this concise overview of geodesic and speed efficiencies, we are now
ready to present the concepts of curvature and the complexity associated with
quantum evolutions.

\section{Curvature and Complexity}

In this section, we will characterize the geometric properties of quantum
evolutions by introducing the concepts of curvature coefficient and the
complexity of the dynamical paths taken by state vectors, which are governed
by nonstationary Hamiltonians.

\subsection{Curvature}

We commence with the notion of the curvature coefficient in quantum evolution.
Following a few mathematical details and a formal definition, we concentrate
on the explicit computation of this coefficient for both higher-dimensional
systems and two-level systems.

\subsubsection{Preliminaries}

In the broadest sense, consider a nonstationary Hamiltonian evolution
characterized by Schr\"{o}dinger's equation $i\hslash\partial_{t}\left\vert
\psi\left(  t\right)  \right\rangle =\mathrm{H}\left(  t\right)  \left\vert
\psi\left(  t\right)  \right\rangle $, where $\left\vert \psi\left(  t\right)
\right\rangle $ is an element of an arbitrary $N$-dimensional complex Hilbert
space $\mathcal{H}_{N}$. Generally, the normalized state vector $\left\vert
\psi\left(  t\right)  \right\rangle $ satisfies the condition $\left\langle
\psi\left(  t\right)  \left\vert \dot{\psi}\left(  t\right)  \right.
\right\rangle =(-i/\hslash)\left\langle \psi\left(  t\right)  \left\vert
\mathrm{H}\left(  t\right)  \right\vert \psi\left(  t\right)  \right\rangle
\neq0$. For the state $\left\vert \psi\left(  t\right)  \right\rangle $, we
define the parallel transported unit state vector $\left\vert \Psi\left(
t\right)  \right\rangle \overset{\text{def}}{=}e^{i\beta\left(  t\right)
}\left\vert \psi\left(  t\right)  \right\rangle $ where the phase
$\beta\left(  t\right)  $ is determined such that $\left\langle \Psi\left(
t\right)  \left\vert \dot{\Psi}\left(  t\right)  \right.  \right\rangle =0$.
Note that $i\hslash\left\vert \dot{\Psi}\left(  t\right)  \right\rangle
=\left[  \mathrm{H}\left(  t\right)  -\hslash\dot{\beta}\left(  t\right)
\right]  \left\vert \Psi\left(  t\right)  \right\rangle $. Consequently, the
relation $\left\langle \Psi\left(  t\right)  \left\vert \dot{\Psi}\left(
t\right)  \right.  \right\rangle =0$ is equivalent to setting $\beta\left(
t\right)  \overset{\text{def}}{=}(1/\hslash)\int_{0}^{t}\left\langle
\psi\left(  t^{\prime}\right)  \left\vert \mathrm{H}\left(  t^{\prime}\right)
\right\vert \psi\left(  t^{\prime}\right)  \right\rangle dt^{\prime}$. For
this reason, $\left\vert \Psi\left(  t\right)  \right\rangle $ simplifies to%
\begin{equation}
\left\vert \Psi\left(  t\right)  \right\rangle =e^{(i/\hslash)\int_{0}%
^{t}\left\langle \psi\left(  t^{\prime}\right)  \left\vert \mathrm{H}\left(
t^{\prime}\right)  \right\vert \psi\left(  t^{\prime}\right)  \right\rangle
dt^{\prime}}\left\vert \psi\left(  t\right)  \right\rangle \text{,}
\label{reason}%
\end{equation}
and fulfills the evolution equation $i\hslash\left\vert \dot{\Psi}\left(
t\right)  \right\rangle =\Delta\mathrm{H}\left(  t\right)  \left\vert
\Psi\left(  t\right)  \right\rangle $ with $\Delta\mathrm{H}\left(  t\right)
\overset{\text{def}}{=}\mathrm{H}\left(  t\right)  -\left\langle
\mathrm{H}\left(  t\right)  \right\rangle $. Note that the speed $v(t)$ of
quantum evolution varies when the Hamiltonian is time-varying. Specifically,
$v(t)$ is defined such that $v^{2}\left(  t\right)  =\left\langle \dot{\Psi
}\left(  t\right)  \left\vert \dot{\Psi}\left(  t\right)  \right.
\right\rangle =\left\langle \left(  \Delta\mathrm{H}\left(  t\right)  \right)
^{2}\right\rangle /\hslash^{2}$. For convenience, we define the arc length
$s=s\left(  t\right)  $ in terms of $v\left(  t\right)  $ as $s\left(
t\right)  \overset{\text{def}}{=}\int_{0}^{t}v(t^{\prime})dt^{\prime}$, with
$ds=v(t)dt$ (which, in turn, implies that $\partial_{t}=v(t)\partial_{s}$).
Evidently, $\partial_{t}\overset{\text{def}}{=}\partial/\partial t$ and
$\partial_{s}\overset{\text{def}}{=}\partial/\partial s$. In conclusion, by
presenting the dimensionless operator%
\begin{equation}
\Delta h\left(  t\right)  \overset{\text{def}}{=}\frac{\Delta\mathrm{H}\left(
t\right)  }{\hslash v(t)}=\frac{\Delta\mathrm{H}\left(  t\right)  }%
{\sqrt{\left\langle \left(  \Delta\mathrm{H}\left(  t\right)  \right)
^{2}\right\rangle }}\text{,}%
\end{equation}
the normalized tangent vector $\left\vert T\left(  s\right)  \right\rangle
\overset{\text{def}}{=}\partial_{s}\left\vert \Psi\left(  s\right)
\right\rangle =\left\vert \Psi^{\prime}\left(  s\right)  \right\rangle $
transforms into $\left\vert T\left(  s\right)  \right\rangle =-i\Delta
h\left(  s\right)  \left\vert \Psi\left(  s\right)  \right\rangle $. It is
noted that $\left\langle T\left(  s\right)  \left\vert T\left(  s\right)
\right.  \right\rangle =1$ by design, and furthermore, $\partial
_{s}\left\langle \Delta h(s)\right\rangle =\left\langle \Delta h^{\prime
}(s)\right\rangle $. We can derive $\left\vert T^{\prime}\left(  s\right)
\right\rangle \overset{\text{def}}{=}\partial_{s}\left\vert T\left(  s\right)
\right\rangle $ from the tangent vector $\left\vert T\left(  s\right)
\right\rangle =-i\Delta h\left(  s\right)  \left\vert \Psi\left(  s\right)
\right\rangle $. Indeed, through algebraic manipulation, we arrive at
$\left\vert T^{\prime}\left(  s\right)  \right\rangle =-i\Delta h(s)\left\vert
\Psi^{\prime}\left(  s\right)  \right\rangle -i\Delta h^{\prime}(s)\left\vert
\Psi\left(  s\right)  \right\rangle $ where $\left\langle T^{\prime}\left(
s\right)  \left\vert T^{\prime}\left(  s\right)  \right.  \right\rangle
=\left\langle \left(  \Delta h^{\prime}(s)\right)  ^{2}\right\rangle
+\left\langle \left(  \Delta h(s)\right)  ^{4}\right\rangle
-2i\operatorname{Re}\left[  \left\langle \Delta h^{\prime}(s)\left(  \Delta
h(s)\right)  ^{2}\right\rangle \right]  \neq1$, in most cases. For clarity, we
emphasize that the fourth power in $\left\langle \left(  \Delta h(s)\right)
^{4}\right\rangle $\textbf{ }that appears in the expression for $\left\langle
T^{\prime}\left(  s\right)  \left\vert T^{\prime}\left(  s\right)  \right.
\right\rangle $ originates from the relation $\left\vert \Psi^{\prime}\left(
s\right)  \right\rangle =-i\Delta h\left(  s\right)  \left\vert \Psi\left(
s\right)  \right\rangle $. We are now ready to introduce the curvature
coefficient for quantum evolutions produced by nonstationary Hamiltonians.

\subsubsection{Definition}

After introducing the vectors $\left\vert \Psi\left(  s\right)  \right\rangle
$, $\left\vert T\left(  s\right)  \right\rangle $, and $\left\vert T^{\prime
}\left(  s\right)  \right\rangle $, we can now define the curvature
coefficient as initially presented in Refs.
\cite{alsing24A,alsing24B,alsing24C}, which is $\kappa_{\mathrm{AC}}%
^{2}\left(  s\right)  \overset{\text{def}}{=}\left\langle \tilde{N}_{\ast
}\left(  s\right)  \left\vert \tilde{N}_{\ast}\left(  s\right)  \right.
\right\rangle $. Observe that $\left\vert \tilde{N}_{\ast}\left(  s\right)
\right\rangle \overset{\text{def}}{=}\mathrm{P}^{\left(  \Psi\right)
}\left\vert T^{\prime}\left(  s\right)  \right\rangle $, where the projection
operator $\mathrm{P}^{\left(  \Psi\right)  }$ onto states that are orthogonal
to $\left\vert \Psi\left(  s\right)  \right\rangle $ is defined as
$\mathrm{P}^{\left(  \Psi\right)  }\overset{\text{def}}{=}\mathrm{I}%
-\left\vert \Psi\left(  s\right)  \right\rangle \left\langle \Psi\left(
s\right)  \right\vert $. Here, \textquotedblleft$\mathrm{I}$\textquotedblright%
\ represents the identity operator in $\mathcal{H}_{N}$. The subscript
\textquotedblleft\textrm{AC}\textquotedblright\ refers to Alsing and Cafaro.
It is noteworthy that the curvature coefficient $\kappa_{\mathrm{AC}}%
^{2}\left(  s\right)  =\left\langle \tilde{N}_{\ast}\left(  s\right)
\left\vert \tilde{N}_{\ast}\left(  s\right)  \right.  \right\rangle $ can be
more conveniently expressed as%
\begin{equation}
\kappa_{\mathrm{AC}}^{2}\left(  s\right)  \overset{\text{def}}{=}\left\Vert
\mathrm{D}\left\vert T(s)\right\rangle \right\Vert ^{2}=\left\Vert
\mathrm{D}^{2}\left\vert \Psi\left(  s\right)  \right\rangle \right\Vert
^{2}\text{,} \label{peggio}%
\end{equation}
where $\mathrm{D}\overset{\text{def}}{=}\mathrm{P}^{\left(  \Psi\right)
}d/ds=\left(  \mathrm{I}-\left\vert \Psi\right\rangle \left\langle
\Psi\right\vert \right)  d/ds$ with $\mathrm{D}\left\vert T(s)\right\rangle
\overset{\text{def}}{=}\mathrm{P}^{\left(  \Psi\right)  }\left\vert T^{\prime
}(s)\right\rangle =$ $\left\vert \tilde{N}_{\ast}\left(  s\right)
\right\rangle $ denoting the covariant derivative
\cite{cafarocqg,samuel88,paulPRA23}. According to Eq. (\ref{peggio}), it is
observed that the curvature coefficient $\kappa_{\mathrm{AC}}^{2}\left(
s\right)  $ corresponds to the square of the magnitude of the second covariant
derivative of the state vector $\left\vert \Psi\left(  s\right)  \right\rangle
$ which defines the quantum Schr\"{o}dinger trajectory within the framework of
projective Hilbert space. To clarify, we highlight that
$\vert$%
$\left\vert \tilde{N}_{\ast}\left(  s\right)  \right\rangle $ is a vector that
is neither orthogonal to the vector $\left\vert T\left(  s\right)
\right\rangle $ nor normalized to one. However, despite lacking proper
normalization, $\left\vert \tilde{N}\left(  s\right)  \right\rangle
\overset{\text{def}}{=}\mathrm{P}^{\left(  T\right)  }\mathrm{P}^{\left(
\Psi\right)  }\left\vert T^{\prime}\left(  s\right)  \right\rangle $ is
orthogonal to $\left\vert T\left(  s\right)  \right\rangle $. Ultimately,
$\left\vert N\left(  s\right)  \right\rangle \overset{\text{def}}{=}$
$\left\vert \tilde{N}\left(  s\right)  \right\rangle /\sqrt{\left\langle
\tilde{N}\left(  s\right)  \left\vert \tilde{N}\left(  s\right)  \right.
\right\rangle }$ represents a normalized vector that is also orthogonal to
$\left\vert T\left(  s\right)  \right\rangle $. In summary, the set $\left\{
\left\vert \Psi\left(  s\right)  \right\rangle \text{, }\left\vert T\left(
s\right)  \right\rangle \text{, }\left\vert N\left(  s\right)  \right\rangle
\right\}  $ constitutes the three orthonormal vectors necessary for
characterizing the curvature of a quantum evolution. It is important to note
that our focus is on the three-dimensional complex subspace spanned by
$\left\{  \left\vert \Psi\left(  s\right)  \right\rangle \text{, }\left\vert
T\left(  s\right)  \right\rangle \text{, }\left\vert N\left(  s\right)
\right\rangle \right\}  $, even though $\mathcal{H}_{N}$ can possess an
arbitrary dimension $N$ as a complex space. Nonetheless, our selection aligns
with the classical geometric viewpoint, where the curvature and torsion
coefficients can be interpreted as the lowest and second-lowest members,
respectively, of a family of generalized curvature functions \cite{alvarez19}.

\subsubsection{Calculation technique: General case}

The explicit computation of the time-dependent curvature coefficient
$\kappa_{\mathrm{AC}}^{2}\left(  s\right)  $ in Eqs. (\ref{peggio}) utilizing
the projection operators formalism often presents challenges. This difficulty
arises because, akin to the classical scenario of space curves in $%
\mathbb{R}
^{3}$ \cite{parker77}, there are fundamentally two issues encountered during
the reparametrization of a quantum curve by its arc length $s$. Firstly, we
may find it impossible to compute $s\left(  t\right)  \overset{\text{def}%
}{=}\int_{0}^{t}v(t^{\prime})dt^{\prime}$ in a closed form. Secondly, even if
we succeed in determining $s=s\left(  t\right)  $, we might still face the
challenge of inverting this relationship, thereby preventing us from obtaining
$t=t\left(  s\right)  $, which is necessary to express $\left\vert \Psi\left(
s\right)  \right\rangle \overset{\text{def}}{=}\left\vert \Psi\left(
t(s)\right)  \right\rangle $. To overcome these obstacles, we can reformulate
$\kappa_{\mathrm{AC}}^{2}\left(  s\right)  $ in Eq. (\ref{peggio}) using
expectation values calculated with respect to the state $\left\vert
\Psi\left(  t\right)  \right\rangle $ (or, alternatively, with respect to
$\left\vert \psi\left(  t\right)  \right\rangle $), which can be determined
without relying on the relation $t=t\left(  s\right)  $
\cite{alsing24A,alsing24B}. For simplicity, we will not explicitly mention the
$s$-dependence of the various operators and expectation values in the
subsequent discussion. For instance, $\Delta h\left(  s\right)  $ will simply
be denoted as $\Delta h$. After performing some algebraic manipulations, we
arrive at $\left\vert \tilde{N}_{\ast}\right\rangle =-\left\{  \left[  \left(
\Delta h\right)  ^{2}-\left\langle \left(  \Delta h\right)  ^{2}\right\rangle
\right]  +i\left[  \Delta h^{\prime}-\left\langle \Delta h^{\prime
}\right\rangle \right]  \right\}  \left\vert \Psi\right\rangle $, with $\Delta
h^{\prime}=\partial_{s}\left(  \Delta h\right)  =\left[  \partial_{t}\left(
\Delta h\right)  \right]  /v\left(  t\right)  $. To assess $\kappa
_{\mathrm{AC}}^{2}\left(  s\right)  \overset{\text{def}}{=}\left\langle
\tilde{N}_{\ast}\left(  s\right)  \left\vert \tilde{N}_{\ast}\left(  s\right)
\right.  \right\rangle $, it is beneficial to introduce the Hermitian operator
$\hat{\alpha}_{1}\overset{\text{def}}{=}\left(  \Delta h\right)
^{2}-\left\langle \left(  \Delta h\right)  ^{2}\right\rangle $ along with the
anti-Hermitian operator $\hat{\beta}_{1}\overset{\text{def}}{=}i\left[  \Delta
h^{\prime}-\left\langle \Delta h^{\prime}\right\rangle \right]  $, where it
holds that $\hat{\beta}_{1}^{\dagger}=-\hat{\beta}_{1}$. Consequently,
$\left\vert \tilde{N}_{\ast}\right\rangle =-\left(  \hat{\alpha}_{1}%
+\hat{\beta}_{1}\right)  \left\vert \Psi\right\rangle $ and $\left\langle
\tilde{N}_{\ast}\left(  s\right)  \left\vert \tilde{N}_{\ast}\left(  s\right)
\right.  \right\rangle $ is equivalent to $\left\langle \hat{\alpha}_{1}%
^{2}\right\rangle -\left\langle \hat{\beta}_{1}^{2}\right\rangle +\left\langle
\left[  \hat{\alpha}_{1}\text{, }\hat{\beta}_{1}\right]  \right\rangle $, with
$\left[  \hat{\alpha}_{1}\text{, }\hat{\beta}_{1}\right]  \overset{\text{def}%
}{=}\hat{\alpha}_{1}\hat{\beta}_{1}-\hat{\beta}_{1}\hat{\alpha}_{1}$
representing the quantum commutator of $\hat{\alpha}_{1}$ and $\hat{\beta}%
_{1}$. It is important to note that the expectation value $\left\langle
\left[  \hat{\alpha}_{1}\text{, }\hat{\beta}_{1}\right]  \right\rangle $ is a
real number, as $\left[  \hat{\alpha}_{1}\text{, }\hat{\beta}_{1}\right]  $
represents a Hermitian operator. This arises from the nature of $\hat{\alpha
}_{1}$ and $\hat{\beta}_{1}$ being Hermitian and anti-Hermitian operators,
respectively. By applying the definitions of $\hat{\alpha}_{1}$ and
$\hat{\beta}_{1}$, we derive $\left\langle \hat{\alpha}_{1}^{2}\right\rangle
=\left\langle (\Delta h)^{4}\right\rangle -\left\langle (\Delta h)^{2}%
\right\rangle ^{2}$, $\left\langle \hat{\beta}_{1}^{2}\right\rangle =-\left[
\left\langle (\Delta h^{\prime})^{2}\right\rangle -\left\langle \Delta
h^{\prime}\right\rangle ^{2}\right]  $, and $\left\langle \left[  \hat{\alpha
}_{1}\text{, }\hat{\beta}_{1}\right]  \right\rangle =i\left\langle \left[
(\Delta h)^{2}\text{, }\Delta h^{\prime}\right]  \right\rangle $. It is
noteworthy that, given that $\left[  (\Delta h)^{2}\text{, }\Delta h^{\prime
}\right]  $ is an anti-Hermitian operator, $\left\langle \left[  (\Delta
h)^{2}\text{, }\Delta h^{\prime}\right]  \right\rangle $ is purely imaginary.
For thoroughness, we emphasize that $\left[  (\Delta h)^{2}\text{, }\Delta
h^{\prime}\right]  $ is not typically a null operator. In fact, we have
$\left[  (\Delta h)^{2}\text{, }\Delta h^{\prime}\right]  =\Delta h\left[
\Delta h\text{, }\Delta h^{\prime}\right]  +\left[  \Delta h\text{, }\Delta
h^{\prime}\right]  \Delta h$ where $\left[  \Delta h\text{, }\Delta h^{\prime
}\right]  =\left[  \mathrm{H}\text{, }\mathrm{H}^{\prime}\right]  $.
Consequently, when we focus for instance on nonstationary qubit Hamiltonians
of the type \textrm{H}$\left(  s\right)  \overset{\text{def}}{=}%
\mathbf{h}\left(  s\right)  \cdot\mathbf{\boldsymbol{\sigma}}$, the commutator
$\left[  \mathrm{H}\text{, }\mathrm{H}^{\prime}\right]  =2i(\mathbf{h}%
\times\mathbf{h}^{\prime})\cdot\mathbf{\boldsymbol{\sigma}}$ may be nonzero,
as the vectors $\mathbf{h}$ and $\mathbf{h}^{\prime}$ are generally not
collinear. Ultimately, a computationally efficient expression for the
curvature coefficient $\kappa_{\mathrm{AC}}^{2}\left(  s\right)  $ in Eq.
(\ref{peggio}) within any arbitrary nonstationary context simplifies to%
\begin{equation}
\kappa_{\mathrm{AC}}^{2}\left(  s\right)  =\left\langle (\Delta h)^{4}%
\right\rangle -\left\langle (\Delta h)^{2}\right\rangle ^{2}+\left[
\left\langle (\Delta h^{\prime})^{2}\right\rangle -\left\langle \Delta
h^{\prime}\right\rangle ^{2}\right]  +i\left\langle \left[  (\Delta
h)^{2}\text{, }\Delta h^{\prime}\right]  \right\rangle \text{.}
\label{curvatime}%
\end{equation}
From Eq. (\ref{curvatime}), it is observed that when the Hamiltonian
\textrm{H} remains constant, $\Delta h^{\prime}$ transforms into the null
operator, allowing us to retrieve the stationary limit $\left\langle (\Delta
h)^{4}\right\rangle -\left\langle (\Delta h)^{2}\right\rangle ^{2}$ for the
curvature coefficient $\kappa_{\mathrm{AC}}^{2}\left(  s\right)  $
\cite{alsing24A}. The formulation of $\kappa_{\mathrm{AC}}^{2}\left(
s\right)  $ in Eq. (\ref{curvatime}) is derived through a method that depends
on the computation of expectation values, which necessitate an understanding
of the state vector $\left\vert \psi\left(  t\right)  \right\rangle $ governed
by the time-dependent Schr\"{o}dinger's evolution equation. As discussed in
Ref. \cite{alsing24B}, this expectation-values methodology provides a valuable
statistical interpretation for $\kappa_{\mathrm{AC}}^{2}\left(  s\right)  $.

\subsubsection{Calculation technique: Qubit case}

Notwithstanding its statistical significance, $\kappa_{\mathrm{AC}}^{2}\left(
s\right)  $ in Eq. (\ref{curvatime}) lacks a definitive geometrical
interpretation. This inadequacy prompts an investigation into nonstationary
Hamiltonians and two-level quantum systems, leading to the derivation of a
closed-form expression for the curvature coefficient associated with a curve
delineated by a single-qubit quantum state evolving under an arbitrary
nonstationary Hamiltonian. The curvature coefficient $\kappa_{\mathrm{AC}}%
^{2}$ can be articulated entirely in terms of two distinct real
three-dimensional vectors that possess a clear geometric meaning. In
particular, these vectors are the Bloch vector $\mathbf{a}\left(  t\right)  $
and the magnetic field vector $\mathbf{h}\left(  t\right)  $. While
$\mathbf{a}\left(  t\right)  $ arises from the density operator $\rho\left(
t\right)  =$ $\left\vert \psi\left(  t\right)  \right\rangle \left\langle
\psi\left(  t\right)  \right\vert \overset{\text{def}}{=}\left[
\mathbf{1}+\mathbf{a}\left(  t\right)  \cdot\mathbf{\boldsymbol{\sigma}%
}\right]  /2$, $\mathbf{h}\left(  t\right)  $ defines the nonstationary
Hamiltonian \textrm{H}$\left(  t\right)  \overset{\text{def}}{=}%
\mathbf{h}\left(  t\right)  \cdot\mathbf{\boldsymbol{\sigma}}$. Based on the
comprehensive examination in Ref. \cite{alsing24B}, one obtains%
\begin{equation}
\kappa_{\mathrm{AC}}^{2}\left(  \mathbf{a}\text{, }\mathbf{h}\right)
=4\frac{\left(  \mathbf{a\cdot h}\right)  ^{2}}{\mathbf{h}^{2}-\left(
\mathbf{a\cdot h}\right)  ^{2}}+\left\{  \frac{\left[  \left(  \mathbf{h}%
^{2}\right)  (\mathbf{\dot{h}}^{2})-\left(  \mathbf{h\cdot\dot{h}}\right)
^{2}\right]  -\left[  \left(  \mathbf{a\cdot\dot{h}}\right)  \mathbf{h-}%
\left(  \mathbf{a\cdot h}\right)  \mathbf{\dot{h}}\right]  ^{2}}{\left[
\mathbf{h}^{2}-\left(  \mathbf{a\cdot h}\right)  ^{2}\right]  ^{3}}%
+4\frac{\left(  \mathbf{a\cdot h}\right)  \left[  \mathbf{a\cdot}\left(
\mathbf{h\times\dot{h}}\right)  \right]  }{\left[  \mathbf{h}^{2}-\left(
\mathbf{a\cdot h}\right)  ^{2}\right]  ^{2}}\right\}  \text{.} \label{XXX}%
\end{equation}
The representation of $\kappa_{\mathrm{AC}}^{2}$ in Eq. (\ref{XXX}) is highly
beneficial from a computational perspective for qubit systems and
simultaneously provides a distinct geometric interpretation of the curvature
of quantum evolution, expressed in terms of the Bloch vector $\mathbf{a}%
\left(  t\right)  $ (normalized, with no units) and the magnetic field vector
$\mathbf{h}\left(  t\right)  $(unnormalized, with $\left[  \mathbf{h}\right]
_{\mathrm{MKSA}}=$\textrm{joules}$=\sec.^{-1}$ when letting $\hslash=1$).

Having elucidated the method for calculating the curvature coefficient of
quantum evolution in two-level systems, we can now proceed to describe the
complexity of quantum evolution on the Bloch sphere.

\subsection{Complexity}

In continuation of the research outlined in Refs. \cite{npb25,epjplus25}, we
assess the single-qubit quantum dynamics governed by both stationary and
nonstationary Hamiltonian evolutions over a defined time interval $\left[
t_{A}\text{, }t_{B}\right]  $ by defining the complexity \textrm{C}$\left(
t_{A}\text{, }t_{B}\right)  $ as%
\begin{equation}
\mathrm{C}\left(  t_{A}\text{, }t_{B}\right)  \overset{\text{def}}{=}%
\frac{\mathrm{V}_{\max}\left(  t_{A}\text{, }t_{B}\right)  -\overline
{\mathrm{V}}\left(  t_{A}\text{, }t_{B}\right)  }{\mathrm{V}_{\max}\left(
t_{A}\text{, }t_{B}\right)  }\text{.} \label{QCD}%
\end{equation}
The rationale behind proposing this expression for the complexity
$\mathrm{C}\left(  t_{A}\text{, }t_{B}\right)  $ will be elucidated in the
subsequent paragraphs.

We commence by defining $\overline{\mathrm{V}}\left(  t_{A}\text{, }%
t_{B}\right)  $ and $\mathrm{V}_{\max}\left(  t_{A}\text{, }t_{B}\right)  $ as
stated in Eq. (\ref{QCD}). To articulate the definition of the so-called
\emph{accessed volume} $\overline{\mathrm{V}}\left(  t_{A}\text{, }%
t_{B}\right)  $, we will adopt a schematic approach as follows. If feasible,
analytically integrate the time-dependent Schr\"{o}dinger evolution equation
$i\hslash\partial_{t}\left\vert \psi(t)\right\rangle =\mathrm{H}\left(
t\right)  \left\vert \psi(t)\right\rangle $ and represent the (normalized)
single-qubit state vector $\left\vert \psi(t)\right\rangle $ at any arbitrary
time $t$ in terms of the computational basis state vectors $\left\{
\left\vert 0\right\rangle \text{, }\left\vert 1\right\rangle \right\}  $. If
this is not feasible, proceed with a numerical analysis. We derive $\left\vert
\psi(t)\right\rangle =c_{0}(t)\left\vert 0\right\rangle +c_{1}(t)\left\vert
1\right\rangle $, with $c_{0}(t)$ and $c_{1}(t)$ being recast as
\begin{equation}
c_{0}(t)\overset{\text{def}}{=}\left\langle 0\left\vert \psi(t)\right.
\right\rangle =\left\vert c_{0}(t)\right\vert e^{i\phi_{0}(t)}\text{, and
}c_{1}(t)\overset{\text{def}}{=}\left\langle 1\left\vert \psi(t)\right.
\right\rangle =\left\vert c_{1}(t)\right\vert e^{i\phi_{1}(t)}\text{,}
\label{qa}%
\end{equation}
respectively. Furthermore, it is important to note that $\phi_{0}(t)$ and
$\phi_{1}(t)$ denote the real phases of $c_{0}(t)$ and $c_{1}(t)$,
respectively. Subsequently, by utilizing the complex quantum amplitudes
$c_{0}(t)$ and $c_{1}(t)$ as presented in Eq. (\ref{qa}), reformulate the
state $\left\vert \psi(t)\right\rangle $ into a physically equivalent state
expressed in its standard Bloch sphere representation, which is defined by the
polar angle $\theta\left(  t\right)  \in\left[  0\text{, }\pi\right]  $ and
the azimuthal angle $\varphi\left(  t\right)  \in\left[  0\text{, }%
2\pi\right)  $. With the availability of the temporal variations of the two
spherical angles $\theta\left(  t\right)  $ and $\varphi\left(  t\right)  $,
determine the volume of the parametric region that the quantum-mechanical
system explores during its evolution from $\left\vert \psi(t_{A})\right\rangle
=\left\vert A\right\rangle $ to $\left\vert \psi(t)\right\rangle $. Finally,
compute the temporal-average volume of the parametric region traversed by the
quantum-mechanical system as it evolves from $\left\vert \psi(t_{A}%
)\right\rangle =\left\vert A\right\rangle $ to $\left\vert \psi(t_{B}%
)\right\rangle =\left\vert B\right\rangle $ with $t\in\left[  t_{A}\text{,
}t_{B}\right]  $.

In accordance with this preliminary outline, we shall now explore the details
of the calculation process for $\overline{\mathrm{V}}\left(  t_{A}\text{,
}t_{B}\right)  $. Utilizing Eq. (\ref{qa}), we observe that $\left\vert
\psi(t)\right\rangle =c_{0}(t)\left\vert 0\right\rangle +c_{1}(t)\left\vert
1\right\rangle $ is physically equivalent to the state $\left\vert
c_{0}(t)\right\vert \left\vert 0\right\rangle +\left\vert c_{1}(t)\right\vert
e^{i\left[  \phi_{1}(t)-\phi_{0}(t)\right]  }\left\vert 1\right\rangle $.
Consequently, $\left\vert \psi(t)\right\rangle $ can be reformulated as
\begin{equation}
\left\vert \psi(t)\right\rangle =\cos\left[  \frac{\theta\left(  t\right)
}{2}\right]  \left\vert 0\right\rangle +e^{i\varphi\left(  t\right)  }%
\sin\left[  \frac{\theta\left(  t\right)  }{2}\right]  \left\vert
1\right\rangle \text{.} \label{qa2}%
\end{equation}
From a formal perspective, the polar angle $\theta\left(  t\right)  $ and the
azimuthal angle $\varphi\left(  t\right)  \overset{\text{def}}{=}\phi
_{1}(t)-\phi_{0}(t)=\arg\left[  c_{1}(t)\right]  -\arg\left[  c_{0}(t)\right]
$ in Eq. (\ref{qa2}) can be expressed as%
\begin{equation}
\theta\left(  t\right)  \overset{\text{def}}{=}2\arctan\left(  \frac
{\left\vert c_{1}(t)\right\vert }{\left\vert c_{0}(t)\right\vert }\right)
\text{,} \label{teta}%
\end{equation}
and, under the conditions that $\operatorname{Re}\left[  c_{1}(t)\right]  >0$
and $\operatorname{Re}\left[  c_{0}(t)\right]  >0$,%
\begin{equation}
\varphi\left(  t\right)  \overset{\text{def}}{=}\arctan\left\{  \frac
{\operatorname{Im}\left[  c_{1}(t)\right]  }{\operatorname{Re}\left[
c_{1}(t)\right]  }\right\}  -\arctan\left\{  \frac{\operatorname{Im}\left[
c_{0}(t)\right]  }{\operatorname{Re}\left[  c_{0}(t)\right]  }\right\}
\text{,} \label{fi}%
\end{equation}
respectively.

In general, the functional form for $\varphi\left(  t\right)  $ as presented
in Eq. (\ref{fi}) can take on a more convoluted expression. This complication
arises from the necessity to express the phase $\arg\left(  z\right)  $ of a
complex number $%
\mathbb{C}
\ni z\overset{\text{def}}{=}x+iy=\left\vert z\right\vert e^{i\arg(z)}$ in
terms of the $2$-\textrm{argument arctangent} function \textrm{atan}$2$ as
$\arg(z)=$ \textrm{atan}$2(y$, $x)$. When $x>0$, the function \textrm{atan}%
$2(y$, $x)$ simplifies to $\arctan\left(  y/x\right)  $. For further
mathematical insights regarding \textrm{atan}$2$, we recommend consulting Ref.
\cite{grad00}. Therefore, with $\theta\left(  t\right)  $ and $\varphi\left(
t\right)  $ defined, it can be observed that the unit three-dimensional Bloch
vector $\mathbf{a}\left(  t\right)  $, which corresponds to the state vector
$\left\vert \psi(t)\right\rangle $ as indicated in Eq. (\ref{qa2}), is
expressed as $\mathbf{a}\left(  t\right)  =(\sin\left[  \theta\left(
t\right)  \right]  \cos\left[  \varphi\left(  t\right)  \right]  $,
$\sin\left[  \theta\left(  t\right)  \right]  \sin\left[  \varphi\left(
t\right)  \right]  $, $\cos\left[  \theta\left(  t\right)  \right]  )$. At
present, we can define $\overline{\mathrm{V}}\left(  t_{A}\text{, }%
t_{B}\right)  $. In particular, the accessed volume $\overline{\mathrm{V}%
}\left(  t_{A}\text{, }t_{B}\right)  $, which relates to the quantum evolution
directed by the Hamiltonian \textrm{H}$\left(  t\right)  $ from $\left\vert
\psi(t_{A})\right\rangle =\left\vert A\right\rangle $ to $\left\vert
\psi(t_{B})\right\rangle =\left\vert B\right\rangle $, where $t$ is within the
interval $\left[  t_{A}\text{, }t_{B}\right]  $, is expressed as%
\begin{equation}
\overline{\mathrm{V}}\left(  t_{A}\text{, }t_{B}\right)  \overset{\text{def}%
}{=}\frac{1}{t_{B}-t_{A}}\int_{t_{A}}^{t_{B}}V(t)dt\text{.}
\label{avgcomplexity}%
\end{equation}
From Eq. (\ref{avgcomplexity}), $\overline{\mathrm{V}}\left(  t_{A}\text{,
}t_{B}\right)  $ can be regarded as a mean value of $V(t)$ over the time
interval $\left[  t_{A}\text{, }t_{B}\right]  $. The variable $V(t)$
referenced in Eq. (\ref{avgcomplexity}) represents the instantaneous volume,
which is defined as%
\begin{equation}
V(t)=V(\theta(t)\text{, }\varphi\left(  t\right)  )\overset{\text{def}%
}{=}\mathrm{vol}\left[  \mathcal{D}_{\mathrm{accessed}}\left[  \theta
(t)\text{, }\varphi(t)\right]  \right]  \text{,} \label{local-complexity}%
\end{equation}
with $\mathrm{vol}\left[  \mathcal{D}_{\mathrm{accessed}}\left[
\theta(t)\text{, }\varphi(t)\right]  \right]  $ being defined as%
\begin{equation}
\mathrm{vol}\left[  \mathcal{D}_{\mathrm{accessed}}\left[  \theta(t)\text{,
}\varphi(t)\right]  \right]  \overset{\text{def}}{=}\int\int_{\mathcal{D}%
_{\mathrm{accessed}}\left[  \theta(t)\text{, }\varphi(t)\right]  }%
\sqrt{g_{\mathrm{FS}}\left(  \theta\text{, }\varphi\right)  }d\theta
d\varphi\text{.} \label{q3}%
\end{equation}
We note that $\mathrm{vol}\left[  \mathcal{\cdot}\right]  $ stands for
$\left\vert \mathrm{vol}\left[  \mathcal{\cdot}\right]  \right\vert \geq0$,
given our consideration of these volumes as defined by positive real numerical
values. In Eq. (\ref{q3}), $g_{\mathrm{FS}}\left(  \theta\text{, }%
\varphi\right)  \overset{\text{def}}{=}\sqrt{\sin^{2}(\theta)/16}$ represents
the determinant of the matrix linked to the Fubini-Study infinitesimal line
element $ds_{\mathrm{FS}}^{2}\overset{\text{def}}{=}(1/4)\left[  d\theta
^{2}+\sin^{2}(\theta)d\varphi^{2}\right]  $. Finally, $\mathcal{D}%
_{\mathrm{accessed}}\left[  \theta(t)\text{, }\varphi(t)\right]  $ in Eq.
(\ref{q3}) specifies the parametric region that the quantum-mechanical system
explores during its transition from the initial state $\left\vert \psi
(t_{A})\right\rangle =\left\vert A\right\rangle $ to an intermediate state
$\left\vert \psi(t)\right\rangle $, where $t\in\left[  t_{A}\text{, }%
t_{B}\right]  $. It is expressed as%
\begin{equation}
\mathcal{D}_{\mathrm{accessed}}\left[  \theta(t)\text{, }\varphi(t)\right]
\overset{\text{def}}{=}\left[  \theta\left(  t_{A}\right)  \text{, }%
\theta\left(  t\right)  \right]  \times\left[  \varphi\left(  t_{A}\right)
\text{, }\varphi\left(  t\right)  \right]  \subset\left[  0\text{, }%
\pi\right]  _{\theta}\times\left[  0\text{, }2\pi\right)  _{\varphi}\text{.}
\label{j5B}%
\end{equation}
To enhance computational efficiency, we observe that the instantaneous volume
$V(t)$ in Eq. (\ref{local-complexity}) can be easily reformulated as
$V(t)=\left\vert \left(  \cos\left[  \theta\left(  t_{A}\right)  \right]
-\cos\left[  \theta\left(  t\right)  \right]  \right)  \left(  \varphi
(t)-\varphi(t_{A})\right)  \right\vert /4$, where $\theta\left(  t\right)  $
and $\varphi\left(  t\right)  $ are defined Eqs. (\ref{teta}) and (\ref{fi}),
respectively. In conclusion, assuming that the tilde symbol denotes the
time-average process, the accessed volume $\overline{\mathrm{V}}\left(
t_{A}\text{, }t_{B}\right)  $ in Eq. (\ref{avgcomplexity}) can be expressed as%
\begin{equation}
\overline{\mathrm{V}}\left(  t_{A}\text{, }t_{B}\right)  \overset{\text{def}%
}{=}\widetilde{\mathrm{vol}}\left[  \mathcal{D}_{\mathrm{accessed}}\left[
\theta(t)\text{, }\varphi(t)\right]  \right]  \text{,} \label{cafe1}%
\end{equation}
with $t\in\left[  t_{A}\text{, }t_{B}\right]  $ in Eq. (\ref{cafe1}). Finally,
in accordance with the physical motivations presented in Refs.
\cite{npb25,epjplus25}, the \emph{accessible volume} $\mathrm{V}_{\max}\left(
t_{A}\text{, }t_{B}\right)  $ in Eq. (\ref{QCD}) is defined as%
\begin{equation}
\mathrm{V}_{\max}(t_{A}\text{, }t_{B})\overset{\text{def}}{=}\mathrm{vol}%
\left[  \mathcal{D}_{\text{\textrm{accessible}}}\left(  \theta\text{, }%
\varphi\right)  \right]  =\int\int_{\mathcal{D}_{\text{\textrm{accessible}}%
}\left(  \theta\text{, }\varphi\right)  }\sqrt{g_{\mathrm{FS}}\left(
\theta\text{, }\varphi\right)  }d\theta d\varphi\text{.} \label{j4}%
\end{equation}
The quantity $\mathcal{D}_{\text{\textrm{accessible}}}\left(  \theta\text{,
}\varphi\right)  $ mentioned in Eq. (\ref{j4}) denotes the (local) maximally
accessible two-dimensional parametric region that occurs during the
quantum-mechanical transition from $\left\vert \psi_{A}\left(  \theta
_{A}\text{, }\varphi_{A}\right)  \right\rangle $ to $\left\vert \psi\left(
\theta_{B}\text{, }\varphi_{B}\right)  \right\rangle $ and is defined by%
\begin{equation}
\mathcal{D}_{\text{\textrm{accessible}}}\left(  \theta\text{, }\varphi\right)
\overset{\text{def}}{=}\left\{  \left(  \theta\text{, }\varphi\right)
:\theta_{\min}\leq\theta\leq\theta_{\max}\text{, and }\varphi_{\min}%
\leq\varphi\leq\varphi_{\max}\right\}  \text{.} \label{j5}%
\end{equation}
It is important to note that $\theta_{\min}$, $\theta_{\max}$, $\varphi_{\min
}$, and $\varphi_{\max}$ as presented in Eq. (\ref{j5}) are defined as
\begin{equation}
\theta_{\min}\overset{\text{def}}{=}\underset{t_{A}\leq t\leq t_{B}}{\min
}\theta(t)\text{, }\theta_{\max}\overset{\text{def}}{=}\underset{t_{A}\leq
t\leq t_{B}}{\max}\theta(t)\text{, }\varphi_{\min}\overset{\text{def}%
}{=}\underset{t_{A}\leq t\leq t_{B}}{\min}\varphi(t)\text{, and }\varphi
_{\max}\overset{\text{def}}{=}\underset{t_{A}\leq t\leq t_{B}}{\max}%
\varphi(t)\text{,} \label{minmax}%
\end{equation}
in that order. Additionally, it should be observed that $\mathcal{D}%
_{\text{\textrm{accessed}}}\left(  \theta\text{, }\varphi\right)
\subset\mathcal{D}_{\text{\textrm{accessible}}}\left(  \theta\text{, }%
\varphi\right)  \subset\left[  0,\pi\right]  _{\theta}\times\left[  0\text{,
}2\pi\right)  _{\varphi}$. Lastly, with $\overline{\mathrm{V}}\left(
t_{A}\text{, }t_{B}\right)  $ and $\mathrm{V}_{\max}\left(  t_{A}\text{,
}t_{B}\right)  $ available in Eqs. (\ref{avgcomplexity}) and (\ref{j4}),
respectively, we can formally define our proposed complexity notion
\textrm{C}$\left(  t_{A}\text{, }t_{B}\right)  $ in Eq. (\ref{QCD}).

In general terms, we assess the complexity of a quantum evolution from an
initial state to a final state on the Bloch sphere by considering the
proportion of the non-accessed volume of the sphere that lies within the
accessible volume of the sphere itself. Specifically, when the accessible
volume is predominantly (minimally) explored, the complexity is low (high).
For more details, we refer to Refs. \cite{npb25,epjplus25}.

Before transitioning to the next section, we highlight that in the dynamics of
qubits, the curvature coefficient $\kappa_{\mathrm{AC}}^{2}$ in Eq.
(\ref{XXX}) is solely dependent on the Bloch vector $\mathbf{a}$ and the
magnetic field vector $\mathbf{h}$. To experimentally determine $\mathbf{a}$,
one can conduct quantum measurements of the expectation values of the Pauli
spin operators $\left\langle \sigma_{x}\right\rangle $, $\left\langle
\sigma_{y}\right\rangle $, and $\left\langle \sigma_{z}\right\rangle $ because
$\mathbf{a}\left(  t\right)  =\mathrm{tr}\left[  \rho\left(  t\right)
\mathbf{\boldsymbol{\sigma}}\right]  $ with $\rho\left(  t\right)  =\left[
\mathbf{1+a}\left(  t\right)  \cdot\mathbf{\boldsymbol{\sigma}}\right]  /2$.
Furthermore, from the Bloch equation $\mathbf{\dot{a}=}2\mathbf{h\times a}$,
it is evident that only the component of $\mathbf{h}$ that is orthogonal to
$\mathbf{a}$ influences the trajectory of $\mathbf{a}$. Consequently, by
utilizing $\mathbf{a}$ and its derivative $\mathbf{\dot{a}}$, one can
reconstruct the plane of rotation, thereby determining the axis of rotation
$\mathbf{h}$, up to a global energy shift that does not alter the state on the
Bloch sphere. Regarding our complexity measure \textrm{C} in Eq. (\ref{QCD})
for qubit dynamics, it can essentially be derived from the spherical angles
representing a qubit state on the Bloch sphere. Considering the fact that%
\begin{equation}
\mathbf{a}\left(  t\right)  =\left(  \sin(\theta_{t})\cos(\varphi_{t})\text{,
}\sin(\theta_{t})\sin(\varphi_{t})\text{, }\cos(\theta_{t})\right)  =\left(
\left\langle \sigma_{x}\right\rangle \left(  t\right)  \text{, }\left\langle
\sigma_{y}\right\rangle \left(  t\right)  \text{, }\left\langle \sigma
_{z}\right\rangle \left(  t\right)  \right)  \text{,}%
\end{equation}
one can experimentally determine the polar and azimuthal angles $\theta
_{t}=\theta\left(  t\right)  $ and $\varphi_{t}=\varphi\left(  t\right)  $
from the measurements of the expectation values of the Pauli operators. In
particular, we have%
\begin{equation}
\theta\left(  t\right)  =\arccos\left[  \left\langle \sigma_{z}\right\rangle
\left(  t\right)  \right]  \text{, and }\varphi\left(  t\right)
=\text{\textrm{atan2}}\left(  \left\langle \sigma_{y}\right\rangle \left(
t\right)  \text{, }\left\langle \sigma_{x}\right\rangle \left(  t\right)
\right)  \text{,}%
\end{equation}
where \textrm{atan2}$\left(  \cdot\text{, }\cdot\right)  $ is the previously
mentioned $2$-argument arctangent function with \textrm{atan2}$\left(
\left\langle \sigma_{y}\right\rangle \left(  t\right)  \text{, }\left\langle
\sigma_{x}\right\rangle \left(  t\right)  \right)  =\arctan\left[
\left\langle \sigma_{y}\right\rangle \left(  t\right)  /\left\langle
\sigma_{x}\right\rangle \left(  t\right)  \right]  $ if $\left\langle
\sigma_{x}\right\rangle \left(  t\right)  >0$.

With these final observations, we are now prepared to present the Hamiltonian
model that we intend to investigate.

\section{Hamiltonian Model}

In this section, we present a two-parameter family of time-varying
Hamiltonians, which we intend to analyze from a geometric perspective.
Specifically, we focus on the time-dependent configurations of the magnetic
fields that characterize the Hamiltonians, along with the temporal dynamics of
the phase that dictates the relative phase factor involved in the
decomposition of the evolving state in relation to the computational basis
state vectors.

\subsection{Preliminaries}

In Ref. \cite{uzdin12}, the most thorough Hermitian nonstationary qubit
Hamiltonian $\mathrm{H}\left(  t\right)  $ is constructed in such a manner
that it generates the same motion $\pi\left(  \left\vert \psi\left(  t\right)
\right\rangle \right)  $ within the complex projective Hilbert space $%
\mathbb{C}
P^{1}$ (or, equivalently, on the Bloch sphere $S^{2}\cong%
\mathbb{C}
P^{1}$) as $\left\vert \psi\left(  t\right)  \right\rangle $, where the
projection operator $\pi$ is such that $\pi:\mathcal{H}_{2}^{1}\ni\left\vert
\psi\left(  t\right)  \right\rangle \mapsto\pi\left(  \left\vert \psi\left(
t\right)  \right\rangle \right)  \in%
\mathbb{C}
P^{1}$. In general, it can be shown that $\mathrm{H}\left(  t\right)  $ is
expressible as%
\begin{equation}
\mathrm{H}\left(  t\right)  =iE\left\vert \partial_{t}m(t)\right\rangle
\left\langle m(t)\right\vert -iE\left\vert m(t)\right\rangle \left\langle
\partial_{t}m(t)\right\vert \text{,} \label{oppio}%
\end{equation}
where, for simplicity, we define $\left\vert m(t)\right\rangle =\left\vert
m\right\rangle $, $\left\vert \partial_{t}m(t)\right\rangle =\left\vert
\dot{m}\right\rangle $, $E=1$ and, lastly, $\hslash=1$. The unit state vector
$\left\vert m\right\rangle $ fulfills the relations $\pi\left(  \left\vert
m(t)\right\rangle \right)  =\pi\left(  \left\vert \psi\left(  t\right)
\right\rangle \right)  $ and $i\partial_{t}\left\vert m(t)\right\rangle
=\mathrm{H}\left(  t\right)  \left\vert m(t)\right\rangle $. The condition
$\pi\left(  \left\vert m(t)\right\rangle \right)  =\pi\left(  \left\vert
\psi\left(  t\right)  \right\rangle \right)  $ results in the conclusion that
$\left\vert m(t)\right\rangle =c(t)\left\vert \psi\left(  t\right)
\right\rangle $, where $c(t)$ represents a complex function. By asserting that
$\left\langle m\left\vert m\right.  \right\rangle =1$, we deduce that
$\left\vert c(t)\right\vert =1$. This condition implies, consequently, that
$c(t)=e^{i\phi\left(  t\right)  }$ for a specific real phase $\phi\left(
t\right)  $. Next, by utilizing the parallel transport condition $\left\langle
m\left\vert \dot{m}\right.  \right\rangle =\left\langle \dot{m}\left\vert
m\right.  \right\rangle =0$, the phase $\phi\left(  t\right)  $ is established
as $i\int\left\langle \psi\left\vert \dot{\psi}\right.  \right\rangle dt$. As
a consequence, $\left\vert m(t)\right\rangle =\exp(-\int_{0}^{t}\left\langle
\psi(t^{\prime})\left\vert \partial_{t^{\prime}}\psi(t^{\prime})\right.
\right\rangle dt^{\prime})\left\vert \psi\left(  t\right)  \right\rangle $. It
is crucial to highlight that $\mathrm{H}\left(  t\right)  $ in Eq.
(\ref{oppio}) is fundamentally traceless, as it comprises solely off-diagonal
elements with respect to the orthogonal basis \{$\left\{  \left\vert
m\right\rangle \text{, }\left\vert \partial_{t}m\right\rangle \right\}  $.
Also, $\mathrm{H}\left(  t\right)  $ is a linear combination of traceless
Pauli spin matrices. Lastly, the condition $i\partial_{t}\left\vert
m(t)\right\rangle =\mathrm{H}\left(  t\right)  \left\vert m(t)\right\rangle $
signifies that $\left\vert m(t)\right\rangle $ complies with the
Schr\"{o}dinger evolution equation.

After presenting some essential preliminary information regarding Uzdin's
research in Ref. \cite{uzdin12}, we are now prepared to introduce our proposed
time-dependent Hamiltonian.

\subsection{The Hamiltonian}

To begin, let us consider the normalized state vector $\left\vert \psi\left(
t\right)  \right\rangle $ defined as,%
\begin{equation}
\left\vert \psi\left(  t\right)  \right\rangle \overset{\text{def}}{=}%
\cos\left[  \alpha\left(  t\right)  \right]  \left\vert 0\right\rangle
+e^{i\beta\left(  t\right)  }\sin\left[  \alpha\left(  t\right)  \right]
\left\vert 1\right\rangle \text{,} \label{do1}%
\end{equation}
with $\alpha\left(  t\right)  $ and $\beta\left(  t\right)  $ being two
generally time-dependent real parameters. Form Eq. (\ref{do1}), we note that
$\left\langle \psi\left(  t\right)  \left\vert \dot{\psi}\left(  t\right)
\right.  \right\rangle =i\dot{\beta}\left(  t\right)  \sin^{2}\left[
\alpha\left(  t\right)  \right]  \neq0$. Therefore, let us find $\left\vert
m(t)\right\rangle =e^{-i\phi\left(  t\right)  }$ $\left\vert \psi\left(
t\right)  \right\rangle $, with $\left\langle m\left(  t\right)  \left\vert
\dot{m}\left(  t\right)  \right.  \right\rangle =0$. We observe that
$\left\langle m\left(  t\right)  \left\vert \dot{m}\left(  t\right)  \right.
\right\rangle =0$ if and only if $\ \dot{\phi}\left(  t\right)
=-i\left\langle \psi\left(  t\right)  \left\vert \dot{\psi}\left(  t\right)
\right.  \right\rangle =\dot{\beta}\left(  t\right)  \sin^{2}(\left[
\alpha\left(  t\right)  \right]  $, that is%
\begin{equation}
\phi\left(  t\right)  =\int_{0}^{t}\dot{\beta}\left(  t^{\prime}\right)
\sin^{2}\left[  \alpha\left(  t^{\prime}\right)  \right]  dt^{\prime}\text{.}
\label{do4}%
\end{equation}
Given $\left\vert \psi\left(  t\right)  \right\rangle $ and $\phi\left(
t\right)  $ in Eqs. (\ref{do1}) and (\ref{do4}), respectively, the state
$\left\vert m(t)\right\rangle $ reduces to%
\begin{equation}
\left\vert m(t)\right\rangle =e^{-i\int_{0}^{t}\dot{\beta}\left(  t^{\prime
}\right)  \sin^{2}\left[  \alpha\left(  t^{\prime}\right)  \right]
dt^{\prime}}\left\{  \cos\left[  \alpha\left(  t\right)  \right]  \left\vert
0\right\rangle +e^{i\beta\left(  t\right)  }\sin\left[  \alpha\left(
t\right)  \right]  \left\vert 1\right\rangle \right\}  \text{,} \label{do4b}%
\end{equation}
Setting $\left\vert m\right\rangle \left\langle m\right\vert =(1/2)\left(
\mathbf{1+a}\cdot\mathbf{\boldsymbol{\sigma}}\right)  $, use of Eq.
(\ref{do4b}) leads to the expression of the Bloch vector $\mathbf{a}\left(
t\right)  $ in terms of the real time-dependent parameters $\alpha\left(
t\right)  $ and $\beta\left(  t\right)  $,%
\begin{equation}
\mathbf{a}\overset{\text{def}}{\mathbf{=}}\left(
\begin{array}
[c]{c}%
a_{x}\\
a_{y}\\
a_{z}%
\end{array}
\right)  =\left(
\begin{array}
[c]{c}%
\sin(2\alpha)\cos\left(  \beta\right) \\
\sin(2\alpha)\sin\left(  \beta\right) \\
\cos(2\alpha)
\end{array}
\right)  \text{.} \label{do6}%
\end{equation}
From Eq. (\ref{do6}), we note that $\mathbf{a\cdot a=}a_{x}^{2}+a_{y}%
^{2}+a_{z}^{2}=1$. Lastly, to find the the expression of $\mathrm{H}\left(
t\right)  =i\left\vert \partial_{t}m(t)\right\rangle \left\langle
m(t)\right\vert -i\left\vert m(t)\right\rangle \left\langle \partial
_{t}m(t)\right\vert $ recast as $h_{0}\left(  t\right)  \mathbf{1}%
+\mathbf{h}\left(  t\right)  \cdot\mathbf{\boldsymbol{\sigma}}$, we need to
find the explicit formulae for both $h_{0}\left(  t\right)  $ and the magnetic
field vector $\mathbf{h}\left(  t\right)  $. After some algebraic
manipulations, we arrive at
\begin{align}
\mathrm{H}\left(  t\right)   &  =\left(
\begin{array}
[c]{cc}%
h_{0}+h_{z} & h_{x}-ih_{y}\\
h_{x}+ih_{y} & h_{0}-h_{z}%
\end{array}
\right) \nonumber\\
&  =\left(
\begin{array}
[c]{cc}%
2\dot{\phi}\cos^{2}\left(  \alpha\right)  & e^{-i\beta}\left[  2\dot{\phi}%
\sin\left(  \alpha\right)  \cos\left(  \alpha\right)  -i\dot{\alpha}%
-\dot{\beta}\sin\left(  \alpha\right)  \cos\left(  \alpha\right)  \right] \\
e^{i\beta}\left[  2\dot{\phi}\sin\left(  \alpha\right)  \cos\left(
\alpha\right)  +i\dot{\alpha}-\dot{\beta}\sin\left(  \alpha\right)
\cos(\alpha)\right]  & 2\dot{\phi}\sin^{2}(\alpha)-2\dot{\beta}\sin^{2}\left(
\alpha\right)
\end{array}
\right)  \text{.} \label{dada2}%
\end{align}
From Eq. (\ref{dada2}), we determine that $h_{0}\left(  t\right)  =0$ (as
expected, since the Hamiltonian is traceless). In addition, using Eq.
(\ref{do4}), we note from Eq. (\ref{dada2}) that $\mathbf{h}\left(  t\right)
$ equals%
\begin{equation}
\mathbf{h}\left(  t\right)  \overset{\text{def}}{=}\left(
\begin{array}
[c]{c}%
h_{x}(t)\\
h_{y}(t)\\
h_{z}(t)
\end{array}
\right)  =\left(
\begin{array}
[c]{c}%
-\frac{\dot{\beta}}{2}\cos(2\alpha)\sin(2\alpha)\cos\left(  \beta\right)
-\dot{\alpha}\sin\left(  \beta\right) \\
-\frac{\dot{\beta}}{2}\cos(2\alpha)\sin(2\alpha)\sin\left(  \beta\right)
+\dot{\alpha}\cos\left(  \beta\right) \\
\frac{\dot{\beta}}{2}\sin^{2}(2\alpha)
\end{array}
\right)  \text{.} \label{magno}%
\end{equation}
By employing Eqs. (\ref{do6}) and (\ref{magno}), one can verify through some
simple but laborious algebra that the Bloch vector and the magnetic vector
fulfill the differential equation $\mathbf{\dot{a}=}2\mathbf{h}\times
\mathbf{a}$ \cite{feynman57}. For completeness, we remark that up to constant
factors, this vector differential equation characterizes the Larmor precession
described by the equation\textbf{ }$d\mathbf{m/}dt=\gamma\mathbf{m\times B}$,
with $\mathbf{m}$ being the magnetic moment vector, $\mathbf{B}$ denoting the
external magnetic moment, and $\gamma$ representing the gyromagnetic
ratio\textbf{. }Interestingly, we note that the Bloch vector $\mathbf{a}%
\left(  t\right)  $ in Eq. (\ref{do6}) and the magnetic field $\mathbf{h}%
\left(  t\right)  $ in Eq. (\ref{magno}) are orthogonal since $\mathbf{a}%
\cdot\mathbf{h=}0$ at any time. Observe that $\mathbf{\dot{a}=}2\mathbf{h}%
\times\mathbf{a}$ implies that $\mathbf{\dot{a}\cdot h=}0$. Therefore, since
$\mathbf{a}\cdot\mathbf{h=}0$ implies that $\mathbf{\dot{a}}\cdot
\mathbf{h+\mathbf{a}\cdot\mathbf{\dot{h}}=}0$, we also have
$\mathbf{\mathbf{a}\cdot\mathbf{\dot{h}}=}0$. In summary, we have the
following relations: (i) $\mathbf{\dot{a}=}2\mathbf{h}\times\mathbf{a}$, which
implies $\mathbf{\dot{a}\cdot h=}0$; (ii) $\mathbf{a}\cdot\mathbf{h=}0$, which
implies $\mathbf{\dot{a}}\cdot\mathbf{h+\mathbf{a}\cdot\mathbf{\dot{h}}=}0$;
(iii) $\mathbf{\mathbf{a}\cdot\mathbf{\dot{h}}=}0$, as a consequence of (i)
and (ii). Exploiting these geometric constraints between the Bloch and
magnetic vectors, the curvature coefficient $\kappa_{\mathrm{AC}}^{2}\left(
\mathbf{a}\text{, }\mathbf{h}\right)  $ in Eq. (\ref{XXX}) reduces to%
\begin{equation}
\kappa_{\mathrm{AC}}^{2}\left(  \mathbf{h}\right)  =\frac{\left(
\mathbf{h}^{2}\right)  (\mathbf{\dot{h}}^{2})-\left(  \mathbf{h\cdot\dot{h}%
}\right)  ^{2}}{\mathbf{h}^{6}}\text{.} \label{do20}%
\end{equation}
Lastly, we note that $\left\Vert \mathbf{h}\right\Vert ^{2}=\left(  \dot
{\beta}^{2}/4\right)  \sin^{2}(2\alpha)+\dot{\alpha}^{2}$, we realize that the
magnetic field vector $\mathbf{h}\left(  t\right)  $ in Eq. (\ref{magno}) can
be recast as $\mathbf{h}\left(  t\right)  =\mathcal{R}_{\hat{z}}\left[
\beta\left(  t\right)  \right]  \mathbf{h}_{0}\left(  t\right)  $. More
specifically, assuming $\left(  \dot{\beta}/2\right)  \sin\left(
2\alpha\right)  \geq0$, we can describe $\mathbf{h}\left(  t\right)  $ in a
matrix form as%
\begin{equation}
\left(
\begin{array}
[c]{c}%
h_{x}\\
h_{y}\\
h_{z}%
\end{array}
\right)  =\mathcal{R}_{\hat{z}}\left[  \beta\left(  t\right)  \right]  \left(
\begin{array}
[c]{c}%
-\sqrt{\left\Vert \mathbf{h}\right\Vert ^{2}-\dot{\alpha}^{2}}\cos(2\alpha)\\
\dot{\alpha}\\
+\sqrt{\left\Vert \mathbf{h}\right\Vert ^{2}-\dot{\alpha}^{2}}\cos(2\alpha)
\end{array}
\right)  \text{.} \label{frank}%
\end{equation}
When $\left(  \dot{\beta}/2\right)  \sin\left(  2\alpha\right)  \leq0$, the
signs of the first and third components of $\mathbf{h}_{0}\left(  t\right)  $
in the right-hand-side of Eq. (\ref{frank}) are reversed. Therefore, Eq.
(\ref{frank}) implies that $\mathbf{h}\left(  t\right)  $ can be described in
terms of a rotation $\mathcal{R}_{\hat{z}}\left[  \beta\left(  t\right)
\right]  $ about the $\hat{z}$-axis by a varying angle $\beta\left(  t\right)
$ of a nonstationary magnetic vector $\mathbf{h}_{0}\left(  t\right)  $. In
particular, when $\beta\left(  t\right)  $ remains constant in time with
$\beta\left(  t\right)  =\beta_{0}$, we have $\left\Vert \mathbf{h}\right\Vert
^{2}=\dot{\alpha}^{2}$, $\mathbf{h}_{0}=\left(  0\text{, }\dot{\alpha}\text{,
}0\right)  $, and $\mathbf{h=}$ $\left(  -\dot{\alpha}\sin\left(  \beta
_{0}\right)  \text{, }\dot{\alpha}\cos(\beta_{0})\text{, }0\right)  $. In Fig.
$1$, we illustrate the temporal behavior of the magnetic field vector
projected onto the $z$-axis in the various scenarios specified by
$\beta\left(  t\right)  $ that will be discussed subsequently.

We are now ready for our applications.\begin{figure}[t]
\centering
\includegraphics[width=0.5\textwidth] {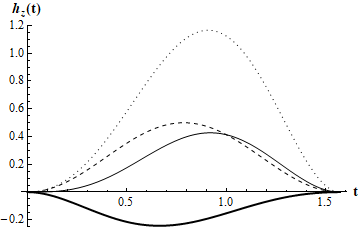}\caption{Temporal behavior of the
magnetic field vector projected onto the $\hat{z}$-axis when the phase
$\beta\left(  t\right)  $ exhibits exponential decay (thick solid line),
quadratic growth (thin solid line), linear growth (dashed line) and, finally,
exponential growth (dotted line). In all plots, we set $\nu_{0}=\omega_{0}=1$
and $0\leq t\leq\pi/2$. Physical units are chosen using $\hslash=1$.}%
\end{figure}

\section{Applications}

In this section, we employ the concepts presented in Sections II and III to
characterize the quantum evolutions outlined in Section IV. Specifically, by
concentrating on the temporal dependence of the previously mentioned phase
$\beta\left(  t\right)  $, we analyze five different scenarios: i) no growth;
ii) linear growth; iii) quadratic growth; iv) exponential growth; v)
exponential decay.

More specifically, we present a comparative analysis of various quantum
evolutions, focusing on the concepts of efficiency, curvature, and complexity
that were introduced earlier. These evolutions are defined by unique magnetic
field configurations $\mathbf{h}\left(  t\right)  $, which are influenced by
the generally time-dependent real parameters $\alpha\left(  t\right)  $ and
$\beta\left(  t\right)  $. Specifically, $\alpha\left(  t\right)  $ dictates
the temporal behavior of the quantum amplitudes of the evolving state
$\left\vert \psi\left(  t\right)  \right\rangle $ in relation to the
computational basis vectors $\left\vert 0\right\rangle $ and $\left\vert
1\right\rangle $, while $\beta\left(  t\right)  $ describes the
quantum-mechanically observable relative phase factor that is included in the
expression of $\left\vert \psi\left(  t\right)  \right\rangle $. We
particularly concentrate on quantum evolutions that transition between
orthogonal initial and final states $\left\vert A\right\rangle
\overset{\text{def}}{=}\left\vert \psi\left(  0\right)  \right\rangle
=\left\vert 0\right\rangle $ and $\left\vert B\right\rangle
\overset{\text{def}}{=}\left\vert \psi\left(  t_{\mathrm{final}}\right)
\right\rangle \simeq\left\vert 1\right\rangle $. Furthermore, to highlight the
changes in motion caused by different time-configurations of the relative
phase factor defined by $\beta\left(  t\right)  $, we will set $\alpha\left(
t\right)  \overset{\text{def}}{=}\omega_{0}t$ with $\omega_{0}\in%
\mathbb{R}
_{+}\backslash\left\{  0\right\}  $, for the entirety of our comparative
analysis. Based on the expression of $\left\vert \psi\left(  t\right)
\right\rangle $, the aforementioned assumption regarding $\alpha\left(
t\right)  $ leads us to take $t_{\mathrm{final}}$ as equal to $\pi
/(2\omega_{0})$. For a schematic overview of selected geometric features of
nonstationary Hamiltonian evolutions on the Bloch sphere, we suggest looking
at Table I.

\subsection{No growth}

In the first example, we assume that $\mathbf{h}\left(  t\right)  $ is
characterized by a phase $\beta\left(  t\right)  $ that does not change in
time. In particular, we set $\beta\left(  t\right)  \overset{\text{def}%
}{=}\beta_{0}\in%
\mathbb{R}
_{+}\backslash\left\{  0\right\}  $ so that $\dot{\beta}=0$. The evolution of
interest occurs from $\left\vert A\right\rangle =\left\vert 0\right\rangle $
to $\left\vert B\right\rangle \simeq\left\vert 1\right\rangle $ in a temporal
interval $t_{\mathrm{final}}=\pi/(2\omega_{0})$, recalling our choice with
$\alpha\left(  t\right)  $ being equal to $\omega_{0}t$. The symbol
\textquotedblleft$\simeq$\textquotedblright\ denotes physical equivalence of
quantum states, modulo unimportant global phase factors.

$\emph{Geodesic}$ $\emph{efficiency}$. From a geodesic efficiency standpoint,
we note that the geodesic distance $s_{0}$ from $\left\vert A\right\rangle $
to $\left\vert B\right\rangle $ is $s_{0}=\pi$. Moreover, the energy
uncertainty $\Delta E\left(  t\right)  =\sqrt{\left\langle \dot{m}\left\vert
\dot{m}\right.  \right\rangle }$ is constant and equals $\omega_{0}$ since
$\left\langle \dot{m}\left\vert \dot{m}\right.  \right\rangle =\dot{\alpha
}^{2}+(1/4)\dot{\beta}^{2}\sin^{2}(2\alpha)$. Therefore, this evolution
happens with unit geodesic efficiency, $\eta_{\mathrm{GE}}=1$, since
$s=s_{0}=\pi$. We stress that the magnetic field vector $\mathbf{h}\left(
t\right)  $ has no longitudinal vector component (i.e., $\mathbf{h}%
_{\parallel}\left(  t\right)  =\mathbf{0}$) since the Bloch vector
$\mathbf{a}\left(  t\right)  $ and $\mathbf{h}\left(  t\right)  $ are
orthogonal for any $0\leq t\leq\pi/(2\omega_{0})$. Indeed, $\mathbf{h}\left(
t\right)  $ is completely transverse since $\mathbf{h}\left(  t\right)
=\mathbf{h}_{\perp}\left(  t\right)  $ and, in addition, $h_{\perp}\left(
t\right)  =\sqrt{\mathbf{h}_{\perp}\cdot\mathbf{h}_{\perp}}=\omega_{0}$ is
constant in time.

\emph{Speed efficiency}. From a speed efficiency standpoint, the evolution
occurs with $\eta_{\mathrm{SE}}(t)=1$ for any $0\leq t\leq\pi/(2\omega_{0})$.
This is justified by construction, given the choice of the Hamiltonian
\textrm{H}$\left(  t\right)  $. In addition, the unit speed efficiency can
also be understood by noticing that magnetic field vector $\mathbf{h}\left(
t\right)  $, as previously mentioned, has no longitudinal vector component
(i.e., $\mathbf{h}_{\parallel}\left(  t\right)  =\mathbf{0}$). It is indeed
the longitudinal vector component of a magnetic field the one that has a
detrimental effect on the motion since its presence limits the fraction of
energy of the systems used for its quantum evolution with a high speed (i.e.,
$\Delta E\left(  t\right)  \leq\left\Vert \mathrm{H}\left(  t\right)
\right\Vert _{\mathrm{SP}}$ when $\mathbf{h}_{\parallel}\left(  t\right)
\neq\mathbf{0}$).

\emph{Curvature}. The curvature coefficient $\kappa_{\mathrm{AC}}^{2}$ of this
quantum evolution is zero. This is is agreement with the fact that
$\eta_{\mathrm{GE}}=1$. From a magnetic field perspective, we have
$\mathbf{h}^{2}=\dot{\alpha}^{2}$, $\mathbf{\dot{h}}^{2}=\ddot{\alpha}^{2}$,
and $\left(  \mathbf{h\cdot\dot{h}}\right)  ^{2}=\dot{\alpha}^{2}\ddot{\alpha
}^{2}$. Therefore, the vanishing of the curvature coefficient is a consequence
of the fact that $\mathbf{h}\left(  t\right)  $ and $\mathbf{\dot{h}}\left(
t\right)  $ are collinear (i.e., $\partial_{t}\hat{h}\left(  t\right)
=\mathbf{0}$, with $\mathbf{h}\left(  t\right)  =h(t)\hat{h}(t)$). In other
words, the magnetic field changes only in intensity, but not in direction.

\emph{Complexity}. Finally, from a complexity viewpoint, we have that
$\theta\left(  t\right)  =2\omega_{0}t$ and $\varphi\left(  t\right)
=\beta_{0}$, with $0\leq\theta\leq\pi$ and $0\leq t\leq\pi/(2\omega_{0})$.
Therefore, a straightforward calculation yields expressions for instantaneous,
accessed, and accessible volumes $V(t)=\omega_{0}t$, $\overline{\mathrm{V}%
}=\pi/4$, and $\mathrm{V}_{\max}=\pi/2$, respectively. Therefore, the
complexity of this quantum evolution reduces to $\mathrm{C}=1/2$%
.\begin{table}[t]
\centering
\begin{tabular}
[c]{c|c|c|c|c}\hline\hline
\textbf{Type of phase}, $\beta\left(  t\right)  $ & \textbf{Geodesic
efficiency}, $\eta_{\mathrm{GE}}$ & \textbf{Speed efficiency}, $\eta
_{\mathrm{SE}}$ & \textbf{Curvature}, $\kappa_{\mathrm{AC}}^{2}$ &
\textbf{Complexity}, \textrm{C}\\\hline
No growth & $1$ & $1$ & $0$ & Constant\\\hline
Linear growth & $<1$ & $1$ & $>0$ & Constant\\\hline
Quadratic growth & $<1$ & $1$ & $>0$ & Constant\\\hline
Exponential growth & $<1$ & $1$ & $>0$ & Non-constant\\\hline
Exponential decay & $<1$ & $1$ & $>0$ & Non-constant\\\hline
\end{tabular}
\caption{Schematic overview of selected geometric features of nonstationary
Hamiltonian evolutions on the Bloch sphere. In particular, for each evolution
defined by a specific time-dependent phase that determines the local phase
factor of the evolving state vector, we examined the geodesic efficiency
$\eta_{\mathrm{GE}}$, speed efficiency $\eta_{\mathrm{SE}}$, curvature
coefficient $\kappa_{\mathrm{AC}}^{2}$, and ultimately, the complexity
$\mathrm{C}$ of the dynamical trajectory on the Bloch sphere that is
determined by the aforementioned state vector. Note that, in general,
$\eta_{\mathrm{GE}}=\eta_{\mathrm{GE}}\left(  \omega_{0}\text{, }\nu
_{0}\right)  $, $\eta_{\mathrm{SE}}=\eta_{\mathrm{SE}}\left(  t\text{; }%
\omega_{0}\text{, }\nu_{0}\right)  $, $\kappa_{\mathrm{AC}}^{2}=\kappa
_{\mathrm{AC}}^{2}\left(  t\text{; }\omega_{0}\text{, }\nu_{0}\right)  $,
and\textrm{ }$\mathrm{C}=\mathrm{C}\left(  \omega_{0}\text{, }\nu_{0}\right)
$. For instance, focusing on global (i.e., non instantaneous) quantities such
as the geodesic efficiency $\eta_{\mathrm{GE}}$ and the complexity
$\mathrm{C}$ and, in addition, setting $\omega_{0}=\nu_{0}=1$, we have:
$\eta_{\mathrm{GE}}^{\mathrm{(No}\text{ \textrm{growth)}}}=1$, $\eta
_{\mathrm{GE}}^{\mathrm{(Linear}\text{ \textrm{growth)}}}\simeq0.94$,
$\eta_{\mathrm{GE}}^{\mathrm{(Quadratic}\text{ \textrm{growth)}}}\simeq0.96$,
$\eta_{\mathrm{GE}}^{\mathrm{(Exponential}\text{ \textrm{growth)}}}\simeq
0.78$, $\eta_{\mathrm{GE}}^{\mathrm{(Exponential}\text{ \textrm{decay)}}%
}\simeq0.98$, $\mathrm{C}^{\mathrm{(No}\text{ \textrm{growth)}}}=0.5$,
$\mathrm{C}^{\mathrm{(Linear}\text{ \textrm{growth)}}}\simeq0.65$,
$\mathrm{C}^{\mathrm{(Quadratic}\text{ \textrm{growth)}}}\simeq0.73$,
$\mathrm{C}^{\mathrm{(Exponential}\text{ \textrm{growth)}}}\simeq0.71$, and
$\mathrm{C}^{\mathrm{(Exponential}\text{ \textrm{decay)}}}\simeq0.59$. For
more details, we refer to the main sections of this manuscript.}%
\end{table}

\subsection{Linear growth}

In the second example, we suppose that $\mathbf{h}\left(  t\right)  $ is
characterized by a phase $\beta\left(  t\right)  $ that grows linearly in
time. More specifically, we put $\beta\left(  t\right)  \overset{\text{def}%
}{=}\nu_{0}t$ with $\nu_{0}\in%
\mathbb{R}
_{+}\backslash\left\{  0\right\}  $ so that $\dot{\beta}=\nu_{0}$. The
evolution of focus happens from $\left\vert A\right\rangle =\left\vert
0\right\rangle $ to $\left\vert B\right\rangle \simeq\left\vert 1\right\rangle
$ in a time interval $t_{\mathrm{final}}=\pi/(2\omega_{0})$ once one recalls
our choice specified by $\alpha\left(  t\right)  =\omega_{0}t$.

$\emph{Geodesic}$ $\emph{efficiency}$. From a geodesic efficiency perspective,
we observe that the geodesic distance $s_{0}$ from $\left\vert A\right\rangle
$ to $\left\vert B\right\rangle $ equals $s_{0}=\pi$. Moreover, the energy
uncertainty $\Delta E\left(  t\right)  =\sqrt{\left\langle \dot{m}\left\vert
\dot{m}\right.  \right\rangle }$ is not constant since $\left\langle \dot
{m}\left\vert \dot{m}\right.  \right\rangle =\dot{\alpha}^{2}+(1/4)\dot{\beta
}^{2}\sin^{2}(2\alpha)$ yields $\Delta E^{2}\left(  t\right)  =\omega_{0}%
^{2}+(1/4)\nu_{0}^{2}\sin^{2}(2\omega_{0}t)$. Therefore, this evolution occurs
with a non unit geodesic efficiency, $\eta_{\mathrm{GE}}<1$, since
$s>s_{0}=\pi$. In general, the quantity $\eta_{\mathrm{GE}}$ is a function of
the parameters $\omega_{0}$ and $\nu_{0}$ and, in particular, can be
numerically estimated. For instance, for $\omega_{0}=\nu_{0}=1$,
$s\simeq3.33\geq\pi=s_{0}$. Similarly to the previous example, the magnetic
field vector $\mathbf{h}\left(  t\right)  $ has no longitudinal vector
component and, moreover, is completely transverse since $\mathbf{h}\left(
t\right)  =\mathbf{h}_{\perp}\left(  t\right)  $. Unlike the previous case,
however, $h_{\perp}\left(  t\right)  =\sqrt{\mathbf{h}_{\perp}\cdot
\mathbf{h}_{\perp}}$ is not constant in time since $h_{\perp}\left(  t\right)
=\sqrt{\omega_{0}^{2}+(1/4)\nu_{0}^{2}\sin^{2}(2\omega_{0}t)}$.

\emph{Speed efficiency}. From a speed efficiency viewpoint, the evolution
happens with $\eta_{\mathrm{SE}}(t)=1$ for any $0\leq t\leq\pi/(2\omega_{0})$.
Like in the first example, this happens thanks to the functional form of the
Hamiltonian \textrm{H}$\left(  t\right)  $. In addition, the unit speed
efficiency can also be explained in terms of the vanishing longitudinal vector
component (i.e., $\mathbf{h}_{\parallel}\left(  t\right)  =\mathbf{0}$) of the
magnetic field vector $\mathbf{h}\left(  t\right)  $.

\emph{Curvature}. The curvature coefficient $\kappa_{\mathrm{AC}}^{2}$ of this
quantum evolution is not equal to zero. This is is agreement with the fact
that $\eta_{\mathrm{GE}}<1$. After some algebra, $\kappa_{\mathrm{AC}}^{2}$ in
Eq. (\ref{XXX}) reduces to%
\begin{equation}
\left[  \kappa_{\mathrm{AC}}^{2}\left(  t\text{; }\omega_{0}\text{, }\nu
_{0}\right)  \right]  _{\mathrm{Example}\text{-\textrm{2}}}=\frac
{\mathbf{h}^{2}\left(  t\text{; }\omega_{0}\text{, }\nu_{0}\right)
\mathbf{\dot{h}}^{2}\left(  t\text{; }\omega_{0}\text{, }\nu_{0}\right)
-\left[  \mathbf{h}\left(  t\text{; }\omega_{0}\text{, }\nu_{0}\right)
\mathbf{\cdot\dot{h}}\left(  t\text{; }\omega_{0}\text{, }\nu_{0}\right)
\right]  ^{2}}{\mathbf{h}^{6}\left(  t\text{; }\omega_{0}\text{, }\nu
_{0}\right)  }\text{,} \label{wish1}%
\end{equation}
where $\mathbf{h}^{2}$, $\mathbf{\dot{h}}^{2}$, and $\left(  \mathbf{h\cdot
\dot{h}}\right)  ^{2}$ in Eq. (\ref{wish1}) are given by%
\begin{align}
&  \mathbf{h}^{2}\overset{\text{def}}{=}\frac{1}{8}\nu_{0}^{2}+\omega_{0}%
^{2}-\frac{1}{8}\nu_{0}^{2}\cos\left(  4\omega_{0}t\right)  \text{,
}\nonumber\\
&  \mathbf{\dot{h}}^{2}\overset{\text{def}}{=}\frac{1}{32}\nu_{0}^{4}-\frac
{1}{32}\nu_{0}^{4}\cos\left(  8\omega_{0}t\right)  +2\nu_{0}^{2}\omega_{0}%
^{2}+2\nu_{0}^{2}\omega_{0}^{2}\cos\left(  4\omega_{0}t\right)  \text{,}%
\nonumber\\
&  \text{ }\mathbf{h\cdot\dot{h}}\overset{\text{def}}{=}\frac{1}{4}\nu_{0}%
^{2}\omega_{0}\sin\left(  4\omega_{0}t\right)  \text{,}%
\end{align}
respectively. As a side note, we observe that the short-time limit of $\left[
\kappa_{\mathrm{AC}}^{2}\left(  t\text{; }\omega_{0}\text{, }\nu_{0}\right)
\right]  _{\mathrm{Example}\text{-\textrm{2}}}$ is expressed as%
\begin{equation}
\left[  \kappa_{\mathrm{AC}}^{2}\left(  t\text{; }\omega_{0}\text{, }\nu
_{0}\right)  \right]  _{\mathrm{Example}\text{-\textrm{2}}}%
\overset{t\rightarrow0}{\simeq}4\left(  \frac{\nu_{0}}{\omega_{0}}\right)
^{2}-8\left(  \frac{\nu_{0}}{\omega_{0}}\right)  ^{2}\left(  \nu_{0}%
^{2}+2\omega_{0}^{2}\right)  t^{2}+\mathcal{O}\left(  t^{3}\right)  \text{,}%
\end{equation}
with $\left[  \kappa_{\mathrm{AC}}^{2}\left(  t\text{; }\omega_{0}\text{, }%
\nu_{0}\right)  \right]  _{\mathrm{Example}\text{-\textrm{2}}}$ commencing at
the nonzero value of $4\left(  \nu_{0}/\omega_{0}\right)  ^{2}$ at $t=0$.
Lastly, the presence of a non-vanishing curvature coefficient is due to the
fact that $\mathbf{h}\left(  t\right)  $ and $\mathbf{\dot{h}}\left(
t\right)  $ are not collinear (i.e., $\partial_{t}\hat{h}\left(  t\right)
\neq\mathbf{0}$, with $\mathbf{h}\left(  t\right)  =h(t)\hat{h}(t)$). In other
terms, unlike what happens in the first example, the magnetic field changes
both in intensity and direction.

\emph{Complexity}. Finally, from a complexity viewpoint, we have that
$\theta\left(  t\right)  =2\omega_{0}t$ and $\varphi\left(  t\right)  =\nu
_{0}t$, with $0\leq\theta\leq\pi$, $0\leq\varphi\leq\left(  \pi/2\right)
\left(  \nu_{0}/\omega_{0}\right)  $, and $0\leq t\leq\pi/(2\omega_{0})$.
Therefore, a straightforward calculation yields expressions for instantaneous,
accessed, and accessible volumes%
\begin{equation}
V(t)=\frac{\nu_{0}}{4}\left[  1-\cos\left(  2\omega_{0}t\right)  \right]
t\text{, }\overline{\mathrm{V}}=\left(  \frac{1}{4\pi}+\frac{\pi}{16}\right)
\frac{\nu_{0}}{\omega_{0}}\text{, and }\mathrm{V}_{\max}=\frac{\pi}{4}%
\frac{\nu_{0}}{\omega_{0}}\text{,}%
\end{equation}
respectively. Therefore, the complexity of this quantum evolution reduces to
\begin{equation}
\mathrm{C}=\frac{3\pi^{2}-4}{4\pi^{2}}\text{.} \label{com2}%
\end{equation}
Interestingly, we note that $\left[  \mathrm{C}\right]  _{\mathrm{Example}%
\text{-\textrm{2}}}$ in Eq. (\ref{com2}) is approximately equal to $0.65$ and
is greater than $\left[  \mathrm{C}\right]  _{\mathrm{Example}%
\text{-\textrm{1}}}=0.5$. It is noteworthy that in the linear growth scenario,
the complexity $\mathrm{C}$ in Eq. (\ref{com2}) reaches a constant value that
is independent of the parameters $\omega_{0}$ and $\nu_{0}$, which arises from
the fact that both $\overline{\mathrm{V}}$ and $\mathrm{V}_{\max}$ are
proportional to the ratio $\left(  \nu_{0}/\omega_{0}\right)  $%
.\begin{figure}[t]
\centering
\includegraphics[width=0.4\textwidth] {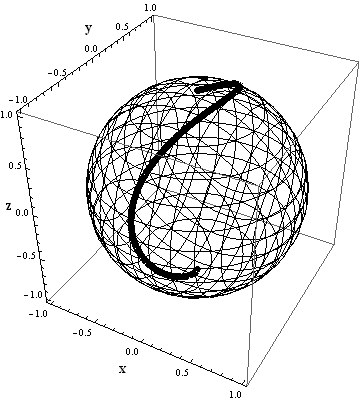}\caption{Illustrative depiction of
the nongeodesic evolution path (thick solid line) on the Bloch sphere
generated by the nonstationary Hamiltonian \textrm{H}$\left(  t\right)  $
associated to a phase $\beta\left(  t\right)  $ that exhibits exponential
growth. The evolution occurs from $\left\vert A\right\rangle
\protect\overset{\text{def}}{=}\left\vert 0\right\rangle $ to $\left\vert
B\right\rangle \protect\overset{\text{def}}{=}\left\vert 1\right\rangle $. For
simplicity, we set $\nu_{0}=\omega_{0}=1$ and $0\leq t\leq\pi/2$. Physical
units are chosen using $\hslash=1$.}%
\end{figure}

\subsection{Quadratic growth}

In the third example, we assume that $\mathbf{h}\left(  t\right)  $ is
specified by a phase $\beta\left(  t\right)  $ that grows quadratically in
time. More precisely, we set $\beta\left(  t\right)  \overset{\text{def}%
}{=}(1/2)\nu_{0}^{2}t^{2}$ with $\nu_{0}\in%
\mathbb{R}
_{+}\backslash\left\{  0\right\}  $ so that $\dot{\beta}=\nu_{0}^{2}t$.
Recalling that $\alpha\left(  t\right)  =\omega_{0}t$, the evolution we study
occurs from $\left\vert A\right\rangle =\left\vert 0\right\rangle $ to
$\left\vert B\right\rangle \simeq\left\vert 1\right\rangle $ in a time
interval $t_{\mathrm{final}}=\pi/(2\omega_{0})$.

$\emph{Geodesic}$ $\emph{efficiency}$. From a geodesic efficiency standpoint,
we note that the geodesic distance $s_{0}$ from $\left\vert A\right\rangle $
to $\left\vert B\right\rangle $ is equal to $s_{0}=\pi$. Furthermore, the
energy uncertainty $\Delta E\left(  t\right)  =\sqrt{\left\langle \dot
{m}\left\vert \dot{m}\right.  \right\rangle }$ changes in time because
$\left\langle \dot{m}\left\vert \dot{m}\right.  \right\rangle =\dot{\alpha
}^{2}+(1/4)\dot{\beta}^{2}\sin^{2}(2\alpha)$ leads to $\Delta E^{2}\left(
t\right)  =\omega_{0}^{2}+(1/4)\nu_{0}^{4}t^{2}\sin^{2}(2\omega_{0}t)$.
Consequently, the quantum evolution in this third scenario happens with a
geodesic efficiency $\eta_{\mathrm{GE}}<1$ because $s>s_{0}=\pi$. Similarly to
the second scenario, the term $\eta_{\mathrm{GE}}$ depends on the two
parameters $\omega_{0}$ and $\nu_{0}$. In this case there is no simple closed
form (i.e., analytical) expression for $\eta_{\mathrm{GE}}$. However, it can
be numerically estimated for a given choice of $\omega_{0}$ and $\nu_{0}$. For
example, setting $\omega_{0}=\nu_{0}=1$, we find $s\simeq3.27\geq\pi=s_{0}$.
In analogy to the first two examples, the magnetic field vector $\mathbf{h}%
\left(  t\right)  $ possesses no longitudinal vector component and, in
addition, is fully transverse since $\mathbf{h}\left(  t\right)
=\mathbf{h}_{\perp}\left(  t\right)  $. Similarly to the second example,
$h_{\perp}\left(  t\right)  =\sqrt{\mathbf{h}_{\perp}\cdot\mathbf{h}_{\perp}}$
is time-varying. In particular, we have $h_{\perp}\left(  t\right)
=\sqrt{\omega_{0}^{2}+(1/4)\nu_{0}^{4}t^{2}\sin^{2}(2\omega_{0}t)}$.

\emph{Speed efficiency}. From the perspective of speed efficiency, the
evolution occurs with $\eta_{\mathrm{SE}}(t)=1$ for any $0\leq t\leq
\pi/(2\omega_{0})$. Similar to the first two examples, this is due to the
functional structure of the Hamiltonian \textrm{H}$\left(  t\right)  $.
Furthermore, the unit speed efficiency can also be elucidated by the absence
of the longitudinal vector component (i.e., $\mathbf{h}_{\parallel}\left(
t\right)  =\mathbf{0}$) of the magnetic field vector $\mathbf{h}\left(
t\right)  $.

\emph{Curvature}. In analogy to what happens in the second application, the
curvature coefficient $\kappa_{\mathrm{AC}}^{2}$ of this third quantum
evolution does not vanish. This agrees with the inequality $\eta_{\mathrm{GE}%
}<1$. After some algebraic manipulations, $\kappa_{\mathrm{AC}}^{2}$ in Eq.
(\ref{XXX}) reduces to%
\begin{equation}
\left[  \kappa_{\mathrm{AC}}^{2}\left(  t\text{; }\omega_{0}\text{, }\nu
_{0}\right)  \right]  _{\mathrm{Example}\text{-\textrm{3}}}=\frac
{\mathbf{h}^{2}\left(  t\text{; }\omega_{0}\text{, }\nu_{0}\right)
\mathbf{\dot{h}}^{2}\left(  t\text{; }\omega_{0}\text{, }\nu_{0}\right)
-\left[  \mathbf{h}\left(  t\text{; }\omega_{0}\text{, }\nu_{0}\right)
\mathbf{\cdot\dot{h}}\left(  t\text{; }\omega_{0}\text{, }\nu_{0}\right)
\right]  ^{2}}{\mathbf{h}^{6}\left(  t\text{; }\omega_{0}\text{, }\nu
_{0}\right)  }\text{,} \label{wish2}%
\end{equation}
where $\mathbf{h}^{2}$, $\mathbf{\dot{h}}^{2}$, and $\left(  \mathbf{h\cdot
\dot{h}}\right)  ^{2}$ in Eq. (\ref{wish2}) are given by%
\begin{align}
&  \mathbf{h}^{2}\overset{\text{def}}{=}\omega_{0}^{2}+\frac{1}{4}\nu_{0}%
^{4}t^{2}\sin^{2}(2\omega_{0}t)\text{, }\nonumber\\
&  \mathbf{\dot{h}}^{2}\overset{\text{def}}{=}\frac{1}{16}\nu_{0}^{8}t^{4}%
\sin^{2}(4\omega_{0}t)+\frac{1}{4}\nu_{0}^{4}\sin^{2}(2\omega_{0}t)+2\nu
_{0}^{4}\omega_{0}^{2}t^{2}\left[  1+\cos(4\omega_{0}t)\right]  +\nu_{0}%
^{4}\omega_{0}t\sin(4\omega_{0}t)\text{,}\nonumber\\
&  \mathbf{h\cdot\dot{h}}\overset{\text{def}}{=}\frac{1}{4}\nu_{0}^{4}t\left[
\omega_{0}t\sin(4\omega_{0}t)+\sin^{2}(2\omega_{0}t)\right]  \text{, }%
\end{align}
respectively. As a supplementary remark, we note that the short-time limit of
$\left[  \kappa_{\mathrm{AC}}^{2}\left(  t\text{; }\omega_{0}\text{, }\nu
_{0}\right)  \right]  _{\mathrm{Example}\text{-\textrm{3}}}$ is represented as%
\begin{equation}
\left[  \kappa_{\mathrm{AC}}^{2}\left(  t\text{; }\omega_{0}\text{, }\nu
_{0}\right)  \right]  _{\mathrm{Example}\text{-\textrm{3}}}%
\overset{t\rightarrow0}{\simeq}9\nu_{0}^{2}\left(  \frac{\nu_{0}}{\omega_{0}%
}\right)  ^{2}t^{2}-28\nu_{0}^{4}t^{4}+\mathcal{O}\left(  t^{5}\right)
\text{,}%
\end{equation}
where $\left[  \kappa_{\mathrm{AC}}^{2}\left(  t\text{; }\omega_{0}\text{,
}\nu_{0}\right)  \right]  _{\mathrm{Example}\text{-\textrm{3}}}$ starts at a
value of zero when $t=0$. Finally, the existence of a non-zero curvature
coefficient arises from the fact that $\mathbf{h}\left(  t\right)  $ and
$\mathbf{\dot{h}}\left(  t\right)  $ are not collinear (i.e., $\partial
_{t}\hat{h}\left(  t\right)  \neq\mathbf{0}$, with $\mathbf{h}\left(
t\right)  =h(t)\hat{h}(t)$). In other words, contrary to the situation that
occurs in the first example, the magnetic field varies in both magnitude and direction.

\emph{Complexity}. Finally, from a complexity perspective, we observe that
$\theta\left(  t\right)  =2\omega_{0}t$ and $\varphi\left(  t\right)
=(1/2)\nu_{0}^{2}t^{2}$, with $0\leq\theta\leq\pi$, $0\leq\varphi
\leq(1/2)\left[  \left(  \pi/2\right)  \left(  \nu_{0}/\omega_{0}\right)
\right]  ^{2}$, and $0\leq t\leq\pi/(2\omega_{0})$. Therefore, a
straightforward computation leads to expressions for the instantaneous,
accessed, and accessible volumes given by%
\begin{equation}
V(t)=\frac{\nu_{0}^{2}}{8}\left[  1-\cos\left(  2\omega_{0}t\right)  \right]
t^{2}\text{, }\overline{\mathrm{V}}=\left(  \frac{1}{16}+\frac{\pi^{2}}%
{96}\right)  \left(  \frac{\nu_{0}}{\omega_{0}}\right)  ^{2}\text{, and
}\mathrm{V}_{\max}=\frac{\pi^{2}}{16}\left(  \frac{\nu_{0}}{\omega_{0}%
}\right)  ^{2}\text{,}%
\end{equation}
respectively. Lastly, the complexity of this quantum evolution becomes
\begin{equation}
\mathrm{C}=\frac{5\pi^{2}-6}{6\pi^{2}}\text{.} \label{com3}%
\end{equation}
Intriguingly, we emphasize that $\left[  \mathrm{C}\right]  _{\mathrm{Example}%
\text{-\textrm{3}}}$ in Eq. (\ref{com3}) is approximately equal to $0.73$ and
is greater than both $\left[  \mathrm{C}\right]  _{\mathrm{Example}%
\text{-\textrm{2}}}\simeq0.65$ and $\left[  \mathrm{C}\right]
_{\mathrm{Example}\text{-\textrm{1}}}=0.5$. Importantly, the observation that
in the quadratic growth scenario, the complexity $\mathrm{C}$ in Eq.
(\ref{com3}) attains a constant value that is independent of the parameters
$\omega_{0}$ and $\nu_{0}$ stems from the fact that both $\overline
{\mathrm{V}}$ and $\mathrm{V}_{\max}$ are proportional to the ratio $\left(
\nu_{0}/\omega_{0}\right)  ^{2}$.

\begin{figure}[t]
\centering
\includegraphics[width=0.5\textwidth] {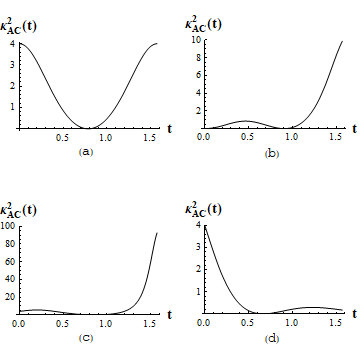}\caption{Temporal behavior of the
curvature coefficient $\kappa_{\mathrm{AC}}^{2}(t)$ of the quantum evolution
when the phase $\beta\left(  t\right)  $ exhibits linear growth (a), quadratic
growth (b), exponential growth (c) and, finally, exponential decay (d). In all
plots, we set $\nu_{0}=\omega_{0}=1$ and $0\leq t\leq\pi/2$. Physical units
are chosen using $\hslash=1$.}%
\end{figure}

\subsection{Exponential growth}

In the fourth example, we suppose that $\mathbf{h}\left(  t\right)  $ is
described by a phase $\beta\left(  t\right)  $ that grows exponentially in
time. Specifically, we put $\beta\left(  t\right)  \overset{\text{def}%
}{=}e^{\nu_{0}t}$ with $\nu_{0}\in%
\mathbb{R}
_{+}\backslash\left\{  0\right\}  $ so that $\dot{\beta}=\nu_{0}e^{\nu_{0}t}$.
Keeping in mind that $\alpha\left(  t\right)  =\omega_{0}t$, the evolution we
investigate happens from $\left\vert A\right\rangle =\left\vert 0\right\rangle
$ to $\left\vert B\right\rangle \simeq\left\vert 1\right\rangle $ in a
temporal interval $t_{\mathrm{final}}=\pi/(2\omega_{0})$. In Figure $2$, we
depict the non-geodesic evolution trajectory on the Bloch sphere produced by
the nonstationary Hamiltonian, which is linked to a phase that demonstrates
exponential growth.

$\emph{Geodesic}$ $\emph{efficiency}$. From a geodesic efficiency perspective,
we observe that the geodesic distance $s_{0}$ from $\left\vert A\right\rangle
$ to $\left\vert B\right\rangle $ equals $s_{0}=\pi$. Moreover, the energy
uncertainty $\Delta E\left(  t\right)  =\sqrt{\left\langle \dot{m}\left\vert
\dot{m}\right.  \right\rangle }$ varies in time since $\left\langle \dot
{m}\left\vert \dot{m}\right.  \right\rangle =\dot{\alpha}^{2}+(1/4)\dot{\beta
}^{2}\sin^{2}(2\alpha)$ yields $\Delta E^{2}\left(  t\right)  =\omega_{0}%
^{2}+(1/4)\nu_{0}^{2}e^{2\nu_{0}t}\sin^{2}(2\omega_{0}t)$. As a consequence,
the quantum evolution in this fourth case exhibits a geodesic efficiency
$\eta_{\mathrm{GE}}<1$ since $s>s_{0}=\pi$. In analogy to the last two
examples, the efficiency $\eta_{\mathrm{GE}}$ is a function of $\omega_{0}$
and $\nu_{0}$. Although there is no useful closed form expression for
$\eta_{\mathrm{GE}}$, one can numerically evaluate it for a suitable choice of
values for $\omega_{0}$ and $\nu_{0}$. For instance, letting $\omega_{0}%
=\nu_{0}=1$, we find $s\simeq4.04\geq\pi=s_{0}$. Analogously to the first
three examples, the magnetic field vector $\mathbf{h}\left(  t\right)  $ is
completely transverse (i.e., $\mathbf{h}\left(  t\right)  =\mathbf{h}_{\perp
}\left(  t\right)  $) and exhibits no longitudinal vector component (i.e.,
$\mathbf{h}_{\parallel}=\mathbf{0}$). In analogy to the second and third
examples, $h_{\perp}\left(  t\right)  =\sqrt{\mathbf{h}_{\perp}\cdot
\mathbf{h}_{\perp}}$ is nonstationary. In particular, we have $h_{\perp
}\left(  t\right)  =\sqrt{\omega_{0}^{2}+(1/4)\nu_{0}^{2}e^{2\nu_{0}t}\sin
^{2}(2\omega_{0}t)}$.

\emph{Speed efficiency}. From the viewpoint of speed efficiency, the evolution
is characterized by $\eta_{\mathrm{SE}}(t)=1$ for any $0\leq t\leq\pi
/(2\omega_{0})$. In a manner akin to the first three examples, this phenomenon
arises from the functional structure of the Hamiltonian \textrm{H}$\left(
t\right)  $. Additionally, the unit speed efficiency can be further clarified
by the lack of the longitudinal vector component (i.e., $\mathbf{h}%
_{\parallel}\left(  t\right)  =\mathbf{0}$) of the magnetic field vector
$\mathbf{h}\left(  t\right)  $.

\emph{Curvature}. In a manner similar to the events occurring in the second
and third applications, the curvature coefficient $\kappa_{\mathrm{AC}}^{2}$
associated with this fourth quantum evolution does not become zero. This is
consistent with the inequality $\eta_{\mathrm{GE}}<1$. Following a series of
algebraic manipulations, $\kappa_{\mathrm{AC}}^{2}$ in Eq. (\ref{XXX})
simplifies to%
\begin{equation}
\left[  \kappa_{\mathrm{AC}}^{2}\left(  t\text{; }\omega_{0}\text{, }\nu
_{0}\right)  \right]  _{\mathrm{Example}\text{-\textrm{4}}}=\frac
{\mathbf{h}^{2}\left(  t\text{; }\omega_{0}\text{, }\nu_{0}\right)
\mathbf{\dot{h}}^{2}\left(  t\text{; }\omega_{0}\text{, }\nu_{0}\right)
-\left[  \mathbf{h}\left(  t\text{; }\omega_{0}\text{, }\nu_{0}\right)
\mathbf{\cdot\dot{h}}\left(  t\text{; }\omega_{0}\text{, }\nu_{0}\right)
\right]  ^{2}}{\mathbf{h}^{6}\left(  t\text{; }\omega_{0}\text{, }\nu
_{0}\right)  }\text{,} \label{wish3}%
\end{equation}
with $\mathbf{h}^{2}$, $\mathbf{\dot{h}}^{2}$, and $\left(  \mathbf{h\cdot
\dot{h}}\right)  ^{2}$ in Eq. (\ref{wish3}) being defined as%
\begin{align}
&  \mathbf{h}^{2}\overset{\text{def}}{=}\omega_{0}^{2}+\frac{1}{4}\nu_{0}%
^{2}e^{2\nu_{0}t}\sin^{2}(2\omega_{0}t)\text{,}\nonumber\\
&  \mathbf{\dot{h}}^{2}\overset{\text{def}}{=}\frac{1}{16}\nu_{0}^{2}%
e^{2t\nu_{0}}\left\{  4\nu_{0}^{2}\sin^{2}(2\omega_{0}t)+32\omega_{0}%
^{2}\left[  1+\cos(4\omega_{0}t)\right]  +\nu_{0}^{2}e^{2t\nu_{0}}\sin
^{2}(4\omega_{0}t)+16\nu_{0}\omega_{0}\sin(4\omega_{0}t)\right\}
\text{,}\nonumber\\
&  \mathbf{h\cdot\dot{h}}\overset{\text{def}}{=}\frac{1}{4}\nu_{0}^{2}%
e^{2t\nu_{0}}\left[  \nu_{0}\sin^{2}(2\omega_{0}t)+\omega_{0}\sin(4\omega
_{0}t)\right]  \text{,}%
\end{align}
respectively. As an additional observation, we observe that the short-time
limit of $\left[  \kappa_{\mathrm{AC}}^{2}\left(  t\text{; }\omega_{0}\text{,
}\nu_{0}\right)  \right]  _{\mathrm{Example}\text{-\textrm{4}}}$ is
articulated as%
\begin{equation}
\left[  \kappa_{\mathrm{AC}}^{2}\left(  t\text{; }\omega_{0}\text{, }\nu
_{0}\right)  \right]  _{\mathrm{Example}\text{-\textrm{4}}}%
\overset{t\rightarrow0}{\simeq}4\left(  \frac{\nu_{0}}{\omega_{0}}\right)
^{2}+12\nu_{0}\left(  \frac{\nu_{0}}{\omega_{0}}\right)  ^{2}t+\left(
\frac{\nu_{0}}{\omega_{0}}\right)  ^{2}\left(  9\nu_{0}^{2}-16\omega_{0}%
^{2}\right)  t^{2}+\mathcal{O}\left(  t^{3}\right)  \text{,}%
\end{equation}
where $\left[  \kappa_{\mathrm{AC}}^{2}\left(  t\text{; }\omega_{0}\text{,
}\nu_{0}\right)  \right]  _{\mathrm{Example}\text{-\textrm{4}}}$ initiates at
the nonzero value of $4\left(  \nu_{0}/\omega_{0}\right)  ^{2}$ at $t=0$.
Lastly, the presence of a non-zero curvature coefficient is due to the
non-collinearity of $\mathbf{h}\left(  t\right)  $ and $\mathbf{\dot{h}%
}\left(  t\right)  $ (i.e., $\partial_{t}\hat{h}\left(  t\right)
\neq\mathbf{0}$, with $\mathbf{h}\left(  t\right)  =h(t)\hat{h}(t)$). In
contrast to the scenario presented in the first example (and, in addition,
similarly to the second and third cases), the magnetic field exhibits
variations in both its magnitude and direction.

\emph{Complexity}. Lastly, from a complexity standpoint, we note that
$\theta\left(  t\right)  =2\omega_{0}t$ and $\varphi\left(  t\right)
=e^{\nu_{0}t}$, with $0\leq\theta\leq\pi$, $1\leq\varphi\leq e^{\frac{\pi}%
{2}\frac{\nu_{0}}{\omega_{0}}}$, and $0\leq t\leq\pi/(2\omega_{0})$.
Consequently, a simple calculation results in formulas for the instantaneous,
accessed, and accessible volumes expressed as%
\begin{equation}
V(t)=\frac{\left[  1-\cos\left(  2\omega_{0}t\right)  \right]  \left(
e^{\nu_{0}t}-1\right)  }{4}\text{, }\overline{\mathrm{V}}=\overline
{\mathrm{V}}\left(  \frac{\nu_{0}}{\omega_{0}}\right)  \overset{\text{def}%
}{=}\frac{\left[  \left(  \frac{\nu_{0}}{\omega_{0}}\right)  ^{2}+2\right]
e^{\frac{\pi}{2}\frac{\nu_{0}}{\omega_{0}}}-2}{\pi\left(  \frac{\nu_{0}%
}{\omega_{0}}\right)  \left[  \left(  \frac{\nu_{0}}{\omega_{0}}\right)
^{2}+4\right]  }-\frac{1}{4}\text{, and }\mathrm{V}_{\max}=\mathrm{V}_{\max
}\left(  \frac{\nu_{0}}{\omega_{0}}\right)  \overset{\text{def}}{=}%
\frac{e^{\frac{\pi}{2}\frac{\nu_{0}}{\omega_{0}}}-1}{2}\text{,}%
\end{equation}
respectively. Finally, the complexity of this fourth quantum evolution
becomes
\begin{equation}
\mathrm{C}=\mathrm{C}\left(  \frac{\nu_{0}}{\omega_{0}}\right)
\overset{\text{def}}{=}\frac{\left[  2\pi\left(  \frac{\nu_{0}}{\omega_{0}%
}\right)  ^{3}-4\left(  \frac{\nu_{0}}{\omega_{0}}\right)  ^{2}+8\pi\frac
{\nu_{0}}{\omega_{0}}-8\right]  e^{\frac{\pi}{2}\frac{\nu_{0}}{\omega_{0}}%
}-\left[  \pi\left(  \frac{\nu_{0}}{\omega_{0}}\right)  ^{3}+4\pi\frac{\nu
_{0}}{\omega_{0}}-8\right]  }{\left[  2\pi\left(  \frac{\nu_{0}}{\omega_{0}%
}\right)  ^{3}+8\pi\frac{\nu_{0}}{\omega_{0}}\right]  e^{\frac{\pi}{2}%
\frac{\nu_{0}}{\omega_{0}}}-\left[  2\pi\left(  \frac{\nu_{0}}{\omega_{0}%
}\right)  ^{3}+8\pi\frac{\nu_{0}}{\omega_{0}}\right]  }\text{.} \label{com4}%
\end{equation}
Unlike what we found in the previous three examples, $\left[  \mathrm{C}%
\right]  _{\mathrm{Example}\text{-\textrm{4}}}=\mathrm{C}\left(  \omega
_{0}\text{, }\nu_{0}\right)  $ in Eq. (\ref{com4}) does not assume a constant
value, regardless of any choice of $\omega_{0}$ and $\nu_{0}$. The behavior of
$\mathrm{C}\left(  \omega_{0}\text{, }\nu_{0}\right)  $ is more complicated.
This complication arises from the fact that $\overline{\mathrm{V}}$ and
$\mathrm{V}_{\max}$ demonstrate a dependence on $\omega_{0}$ and $\nu_{0}$
that cannot be encapsulated by a universal scaling factor of the form $\left(
\nu_{0}/\omega_{0}\right)  ^{n}$ with $n\in%
\mathbb{N}
$, for example. Technical details on the derivation of Eq. (\ref{com4}) along
with a discussion of the limiting cases in which $\nu_{0}/\omega_{0}$
approaches zero or infinity are located in Appendix D.

\subsection{Exponential decay}

In this last example, we assume that $\mathbf{h}\left(  t\right)  $ is
characterized by a phase $\beta\left(  t\right)  $ that decays exponentially
in time. In particular, we set $\beta\left(  t\right)  \overset{\text{def}%
}{=}e^{-\nu_{0}t}$ with $\nu_{0}\in%
\mathbb{R}
_{+}\backslash\left\{  0\right\}  $ so that $\dot{\beta}=-\nu_{0}e^{\nu_{0}t}%
$. Given that $\alpha\left(  t\right)  =\omega_{0}t$, the evolution we
consider propagates from $\left\vert A\right\rangle =\left\vert 0\right\rangle
$ to $\left\vert B\right\rangle \simeq\left\vert 1\right\rangle $ in a a time
span specified by $t_{\mathrm{final}}=\pi/(2\omega_{0})$.

$\emph{Geodesic}$ $\emph{efficiency}$. From a geodesic efficiency viewpoint,
we note that the geodesic distance $s_{0}$ from $\left\vert A\right\rangle $
to $\left\vert B\right\rangle $ is $s_{0}=\pi$. Additionally, the energy
uncertainty $\Delta E\left(  t\right)  =\sqrt{\left\langle \dot{m}\left\vert
\dot{m}\right.  \right\rangle }$ does not remain constant in time because
$\left\langle \dot{m}\left\vert \dot{m}\right.  \right\rangle =\dot{\alpha
}^{2}+(1/4)\dot{\beta}^{2}\sin^{2}(2\alpha)$ leads to $\Delta E^{2}\left(
t\right)  =\omega_{0}^{2}+(1/4)\nu_{0}^{2}e^{-2\nu_{0}t}\sin^{2}(2\omega
_{0}t)$. Therefore, the quantum-mechanical evolution in this last scenario
shows a geodesic efficiency $\eta_{\mathrm{GE}}<1$ given that $s>s_{0}=\pi$.
Analogously to the last three examples, the efficiency $\eta_{\mathrm{GE}}$
depends on $\omega_{0}$ and $\nu_{0}$. Again, even though there is not any
convenient closed form formula for $\eta_{\mathrm{GE}}$, it is possible to
numerically calculate it for a given choice of values for $\omega_{0}$ and
$\nu_{0}$. For example, setting $\omega_{0}=\nu_{0}=1$, we obtain
$s\simeq3.19\geq\pi=s_{0}$. In analogy to the first four examples, the
magnetic field vector $\mathbf{h}\left(  t\right)  $ exhibits a fully
transverse vector component (i.e., $\mathbf{h}\left(  t\right)  =\mathbf{h}%
_{\perp}\left(  t\right)  $) and does not possess any longitudinal vector
component (i.e., $\mathbf{h}_{\parallel}=\mathbf{0}$). Furthermore, $h_{\perp
}\left(  t\right)  =\sqrt{\mathbf{h}_{\perp}\cdot\mathbf{h}_{\perp}}$ is
time-varying, with $h_{\perp}\left(  t\right)  =\sqrt{\omega_{0}^{2}%
+(1/4)\nu_{0}^{2}e^{-2\nu_{0}t}\sin^{2}(2\omega_{0}t)}$.

\emph{Speed efficiency}. From the perspective of speed efficiency, the
evolution is defined by $\eta_{\mathrm{SE}}(t)=1$ for any $0\leq t\leq
\pi/(2\omega_{0})$. Similar to the previous four examples, this occurrence
stems from the functional framework of the Hamiltonian \textrm{H}$\left(
t\right)  $. Furthermore, the unit speed efficiency can be elucidated by the
absence of the longitudinal vector component (i.e., $\mathbf{h}_{\parallel
}\left(  t\right)  =\mathbf{0}$) of the magnetic field vector $\mathbf{h}%
\left(  t\right)  $.

\emph{Curvature}. In a manner akin to the occurrences in the second, third,
and fourth applications, the curvature coefficient $\kappa_{\mathrm{AC}}^{2}$
related to this fifth quantum evolution does not reach zero. This aligns with
the inequality $\eta_{\mathrm{GE}}<1$. After performing a sequence of
algebraic calculations, $\kappa_{\mathrm{AC}}^{2}$ in Eq. (\ref{XXX}) is
simplified to%

\begin{equation}
\left[  \kappa_{\mathrm{AC}}^{2}\left(  t\text{; }\omega_{0}\text{, }\nu
_{0}\right)  \right]  _{\mathrm{Example}\text{-\textrm{5}}}=\frac
{\mathbf{h}^{2}\left(  t\text{; }\omega_{0}\text{, }\nu_{0}\right)
\mathbf{\dot{h}}^{2}\left(  t\text{; }\omega_{0}\text{, }\nu_{0}\right)
-\left[  \mathbf{h}\left(  t\text{; }\omega_{0}\text{, }\nu_{0}\right)
\mathbf{\cdot\dot{h}}\left(  t\text{; }\omega_{0}\text{, }\nu_{0}\right)
\right]  ^{2}}{\mathbf{h}^{6}\left(  t\text{; }\omega_{0}\text{, }\nu
_{0}\right)  }\text{,} \label{wish4}%
\end{equation}
where $\mathbf{h}^{2}$, $\mathbf{\dot{h}}^{2}$, and $\left(  \mathbf{h\cdot
\dot{h}}\right)  ^{2}$ in Eq. (\ref{wish4}) are given by%
\begin{align}
&  \mathbf{h}^{2}\overset{\text{def}}{=}\omega_{0}^{2}+\frac{1}{4}\nu_{0}%
^{2}e^{-2\nu_{0}t}\sin^{2}(2\omega_{0}t)\text{,}\nonumber\\
&  \mathbf{\dot{h}}^{2}\overset{\text{def}}{=}\frac{1}{16}\nu_{0}^{2}%
e^{-2\nu_{0}t}\left\{  4\nu_{0}^{2}\sin^{2}(2\omega_{0}t)+32\omega_{0}%
^{2}\left[  1+\cos(4\omega_{0}t)\right]  +\nu_{0}^{2}e^{-2\nu_{0}t}\sin
^{2}(4\omega_{0}t)-16\nu_{0}\omega_{0}\sin\left(  4\omega_{0}t\right)
\right\}  \text{,}\nonumber\\
&  \mathbf{h\cdot\dot{h}}\overset{\text{def}}{=}\frac{1}{4}\nu_{0}^{2}%
e^{-2\nu_{0}t}\left[  \omega_{0}\sin(4\omega_{0}t)-\nu_{0}\sin^{2}(2\omega
_{0}t)\right]  \text{,}%
\end{align}
respectively. As a side note, we observe that the short-time limit of $\left[
\kappa_{\mathrm{AC}}^{2}\left(  t\text{; }\omega_{0}\text{, }\nu_{0}\right)
\right]  _{\mathrm{Example}\text{-\textrm{5}}}$ is represented by%
\begin{equation}
\left[  \kappa_{\mathrm{AC}}^{2}\left(  t\text{; }\omega_{0}\text{, }\nu
_{0}\right)  \right]  _{\mathrm{Example}\text{-\textrm{5}}}%
\overset{t\rightarrow0}{\simeq}4\left(  \frac{\nu_{0}}{\omega_{0}}\right)
^{2}-12\nu_{0}\left(  \frac{\nu_{0}}{\omega_{0}}\right)  ^{2}t+\left(
\frac{\nu_{0}}{\omega_{0}}\right)  ^{2}\left(  9\nu_{0}^{2}-16\omega_{0}%
^{2}\right)  t^{2}+\allowbreak\mathcal{O}\left(  t^{3}\right)  \text{,}%
\end{equation}
where $\left[  \kappa_{\mathrm{AC}}^{2}\left(  t\text{; }\omega_{0}\text{,
}\nu_{0}\right)  \right]  _{\mathrm{Example}\text{-\textrm{5}}}$ begins at the
nonzero value of $4\left(  \nu_{0}/\omega_{0}\right)  ^{2}$ at $t=0$. Finally,
the existence of a non-zero curvature coefficient arises from the
non-collinearity of $\mathbf{h}\left(  t\right)  $ and $\mathbf{\dot{h}%
}\left(  t\right)  $ (i.e., $\partial_{t}\hat{h}\left(  t\right)
\neq\mathbf{0}$, with $\mathbf{h}\left(  t\right)  =h(t)\hat{h}(t)$). Unlike
the situation described in the first example (and similarly to the second,
third, and fourth examples), the magnetic field shows fluctuations in both its
magnitude and direction.

\emph{Complexity}. Finally, from a complexity perspective, we remark that
$\theta\left(  t\right)  =2\omega_{0}t$ and $\varphi\left(  t\right)
=e^{-\nu_{0}t}$, with $0\leq\theta\leq\pi$, $e^{-\frac{\pi}{2}\frac{\nu_{0}%
}{\omega_{0}}}\leq\varphi\leq1$, and $0\leq t\leq\pi/(2\omega_{0})$.
Therefore, a straightforward computation yields expressions for the
instantaneous, accessed, and accessible volumes given by%
\begin{equation}
V(t)=\frac{\left[  1-\cos\left(  2\omega_{0}t\right)  \right]  \left(
1-e^{-\nu_{0}t}\right)  }{4}\text{, }\overline{\mathrm{V}}=\overline
{\mathrm{V}}\left(  \frac{\nu_{0}}{\omega_{0}}\right)  \overset{\text{def}%
}{=}\frac{1}{4}-\frac{2-\left[  \left(  \frac{\nu_{0}}{\omega_{0}}\right)
^{2}+2\right]  e^{-\frac{\pi}{2}\frac{\nu_{0}}{\omega_{0}}}}{\pi\left(
\frac{\nu_{0}}{\omega_{0}}\right)  \left[  \left(  \frac{\nu_{0}}{\omega_{0}%
}\right)  ^{2}+4\right]  }\text{, and }\mathrm{V}_{\max}=\mathrm{V}_{\max
}\left(  \frac{\nu_{0}}{\omega_{0}}\right)  \overset{\text{def}}{=}%
\frac{1-e^{-\frac{\pi}{2}\frac{\nu_{0}}{\omega_{0}}}}{2}\text{,}%
\end{equation}
respectively. Lastly, the complexity of this last quantum evolution reduces
to
\begin{equation}
\mathrm{C}=\mathrm{C}\left(  \frac{\nu_{0}}{\omega_{0}}\right)
\overset{\text{def}}{=}\frac{\left[  \pi\left(  \frac{\nu_{0}}{\omega_{0}%
}\right)  ^{3}+4\pi\left(  \frac{\nu_{0}}{\omega_{0}}\right)  +8\right]
e^{\frac{\pi}{2}\frac{\nu_{0}}{\omega_{0}}}-\left[  2\pi\left(  \frac{\nu_{0}%
}{\omega_{0}}\right)  ^{3}+4\left(  \frac{\nu_{0}}{\omega_{0}}\right)
^{2}+8\pi\left(  \frac{\nu_{0}}{\omega_{0}}\right)  +8\right]  }{\left[
2\pi\left(  \frac{\nu_{0}}{\omega_{0}}\right)  ^{3}+8\pi\left(  \frac{\nu_{0}%
}{\omega_{0}}\right)  \right]  e^{\frac{\pi}{2}\frac{\nu_{0}}{\omega_{0}}%
}-\left[  2\pi\left(  \frac{\nu_{0}}{\omega_{0}}\right)  ^{3}+8\pi\left(
\frac{\nu_{0}}{\omega_{0}}\right)  \right]  }\text{.} \label{com5}%
\end{equation}
Like what we find in the fourth example, $\left[  \mathrm{C}\right]
_{\mathrm{Example}\text{-\textrm{5}}}=\mathrm{C}\left(  \omega_{0}\text{, }%
\nu_{0}\right)  $ in Eq. (\ref{com5}) is not constant in $\omega_{0}$ and
$\nu_{0}$. The parametric behavior of $\mathrm{C}\left(  \omega_{0}\text{,
}\nu_{0}\right)  $ is more intricate. This intricacy arises from the fact that
$\overline{\mathrm{V}}$ and $\mathrm{V}_{\max}$ show a dependence on
$\omega_{0}$ and $\nu_{0}$ that cannot be expressed, for instance, through a
common scaling factor of the form $\left(  \nu_{0}/\omega_{0}\right)  ^{n}$
with $n\in%
\mathbb{N}
$, for example. Technical details on the derivation of Eq. (\ref{com5}) along
with a discussion of the limiting cases in which $\nu_{0}/\omega_{0}$
approaches zero or infinity are located in Appendix A. In Fig. $3$, we show
the temporal behavior of the curvature coefficient of the the various quantum
evolutions considered here. Finally, we present in Fig. $4$ the parametric
characteristics of the complexity of quantum evolution defined by the
nonstationary Hamiltonian, which corresponds to the phase $\beta\left(
t\right)  $ in $\left\vert \psi\left(  t\right)  \right\rangle =\cos\left(
\omega_{0}t\right)  \left\vert 0\right\rangle +e^{i\beta\left(  t\right)
}\sin\left(  \omega_{0}t\right)  \left\vert 1\right\rangle $ (see Eq.
(\ref{do1})) that either increases exponentially or decreases
exponentially.\begin{figure}[t]
\centering
\includegraphics[width=0.5\textwidth] {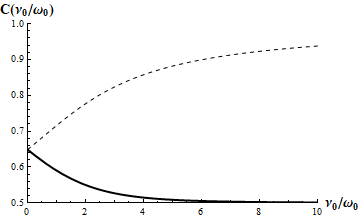}\caption{Behavior of the
complexity of the quantum evolution specified by the nonstationary Hamiltonian
\textrm{H}$\left(  t\right)  $ corresponding to the phase $\beta\left(
t\right)  $ that grows exponentially (dashed line) or decays exponentially
(thick solid line). The complexity \textrm{C}\textsf{\ }is plotted versus the
ratio $\nu_{0}/\omega_{0}$. When $\nu_{0}/\omega_{0}\gg1$ and $\beta\left(
t\right)  $ grows exponentially, the complexity asymptotically approaches its
maximum value $1$. Alternatively, when $\nu_{0}/\omega_{0}\gg1$ and
$\beta\left(  t\right)  $ decays exponentially, the complexity asymptotically
approaches the limiting value of $1/2$, that is to say the complexity value
that characterizes the geodesic evolution on the Bloch sphere between initial
and final states $\left\vert A\right\rangle \protect\overset{\text{def}%
}{=}\left\vert 0\right\rangle $ and $\left\vert B\right\rangle
\protect\overset{\text{def}}{=}\left\vert 1\right\rangle $, respectively.}%
\end{figure}

\section{Final Remarks}

In this paper, we presented a geometric characterization of nonstationary
Hamiltonian evolutions in two-level quantum systems (Eq. (\ref{oppio}) and
Fig. $1$), with special concern on efficiency (Eqs. (\ref{efficiency}) and
(\ref{se1})), curvature (\ref{XXX}), and complexity (\ref{QCD}) of quantum
evolutions (Fig. $2$). In particular, we reported exact analytical expressions
for the curvature of quantum evolutions related to a two-level quantum system
influenced by a variety of time-dependent magnetic field configurations (Eqs.
(\ref{wish1}), (\ref{wish2}), (\ref{wish3}), and (\ref{wish4})). More
specifically, we investigated the dynamics generated by a two-parameter
nonstationary Hermitian Hamiltonian operating at unit speed efficiency. To
deepen our comprehension of the physical implications of the curvature
coefficient, we scrutinized the curvature behavior (Fig. $3$) in connection
with geodesic efficiency, speed efficiency, and the complexity of the quantum
evolution. Our results suggest that, in general, efficient quantum evolutions
tend to exhibit lower complexity than their inefficient counterparts (Table I
and Fig. $4$). Nevertheless, it is important to highlight that complexity is
not solely determined by length. Indeed, longer paths that are sufficiently
curved may reveal a complexity that is less than that of shorter paths
characterized by a lower curvature coefficient. For instance, considering our
second (linear growth, Eq. (\ref{com2})) and third (quadratic growth, Eq.
(\ref{com3})) examples, we have that $\left[  \mathrm{C}\right]
_{\mathrm{Example}\text{-\textrm{2}}}\simeq0.65\leq0.73\simeq\left[
\mathrm{C}\right]  _{\mathrm{Example}\text{-\textrm{3}}}$ does not generally
imply that $\left[  s\left(  \nu_{0}\text{, }\omega_{0}\right)  \right]
_{\mathrm{Example}\text{-\textrm{2}}}\leq\left[  s\left(  \nu_{0}\text{,
}\omega_{0}\right)  \right]  _{\mathrm{Example}\text{-\textrm{3}}}$ (with $s$
defined in Eq. (\ref{efficiency})). Indeed, after some algebra, it happens
that for $\nu_{0}/\omega_{0}\leq2/\pi$, $\left[  s\left(  \nu_{0}\text{,
}\omega_{0}\right)  \right]  _{\mathrm{Example}\text{-\textrm{2}}}\geq\left[
s\left(  \nu_{0}\text{, }\omega_{0}\right)  \right]  _{\mathrm{Example}%
\text{-\textrm{3}}}$. For instance, setting $\nu_{0}=1\left[  \mathrm{MKSA}%
\right]  $ and $\omega_{0}=5\left[  \mathrm{MKSA}\right]  $, one arrives at
$\left[  s\left(  1\text{, }5\right)  \right]  _{\mathrm{Example}%
\text{-\textrm{2}}}\simeq1.575\geq1.571\simeq\left[  s\left(  1\text{,
}5\right)  \right]  _{\mathrm{Example}\text{-\textrm{3}}}$. In addition,
$\left[  \kappa_{\mathrm{AC}}^{2}\left(  t\right)  \right]  _{\mathrm{Example}%
\text{-\textrm{2}}}\geq\left[  \kappa_{\mathrm{AC}}^{2}\left(  t\right)
\right]  _{\mathrm{Example}\text{-\textrm{3}}}$ with $0\leq t\leq\pi/\left(
2\omega_{0}\right)  $ and $\omega_{0}=5\left[  \mathrm{MKSA}\right]  $. These
considerations, arising here from nonstationary Hamiltonian evolutions, align
with those documented in a time-independent Hamiltonian context as referenced
in Ref. \cite{npb25}.

To summarize, our principal findings can be articulated as follows:

\begin{enumerate}
\item[{[i]}] We developed a two-parameter family of nonstationary Hamiltonians
(Eq. (\ref{oppio})) that produce analytically solvable Schr\"{o}dinger
evolution trajectories. Notably, the two parameters (i.e., $\alpha\left(
t\right)  $ and $\beta\left(  t\right)  $) can be interpreted as adjustable
parameters that define the time-varying magnetic field vector, which
characterizes the traceless nonstationary Hamiltonian, or alternatively, the
polar and azimuthal angles (i.e., $\theta\left(  t\right)  $ and
$\varphi\left(  t\right)  $) utilized to parametrize the evolving state
vectors of the Bloch sphere.

\item[{[ii]}] For the first time in the literature (to the best of our
knowledge), we conducted a geometrically-driven comparative analysis of
various dynamical paths on the Bloch sphere that correspond to different
time-varying Hamiltonian models. Each trajectory was defined in terms of
geodesic and speed efficiencies (Eqs. (\ref{efficiency}) and (\ref{se1})),
curvature coefficient (Eq. (\ref{XXX})), and ultimately, complexity of the
corresponding quantum evolution (Eq. (\ref{QCD})).

\item[{[iii]}] We demonstrated the significance of relative phase factors in
quantum dynamics by examining the short-time behavior of the curvature
coefficient across various quantum evolutions, adjusting the time-dependent
phase $\beta\left(  t\right)  $ that appears in the relative phase factor
$e^{i\beta\left(  t\right)  }$ (and, in addition, in the magnetic field vector
$\mathbf{h}$ in Eq. (\ref{magno})) that defines the evolving state vector of
interest (Eqs. (\ref{do1}) and (\ref{do6})). Variations in phases not only
affect the length of the trajectory but also influence the manner in which the
trajectory curves as it connects the initial and final states.

\item[{[iv]}] Changes in lengths and curvature serve as indicators of
alterations in the complexity of the dynamical evolution. Indeed, we
illustrated how these variations are reflected in the accessed and accessible
volumes of the parametric regions on the Bloch sphere (Eqs. (\ref{com2}),
(\ref{com3}), (\ref{com4}), and (\ref{com5})) that arise from the different
nonstationary Hamiltonian models under investigation.
\end{enumerate}

\medskip

In our opinion, this paper represents a notable progress in relation to our
recent works appeared in Refs. \cite{cafaropra25,epjplus25}. Indeed, in
contrast to Ref. \cite{cafaropra25}, we have broadened our investigation to
encompass the curvature of quantum evolutions across various time-dependent
magnetic field configurations, thus covering a diverse range of typical
temporal profiles. Additionally, unlike Ref. \cite{cafaropra25}, we have
examined the temporal behavior of the curvature coefficients and established a
connection to the complexity of the different quantum evolutions. Lastly, in
contrast to Ref. \cite{epjplus25}, we have expanded our analysis to include
the complexity of quantum evolutions characterized by non-stationary
Hamiltonians. To the best of our understanding, most curvature analyses of
quantum evolutions documented in the literature are confined to
time-independent contexts
\cite{laba17,gnatenko22,banks24,gnatenko24,gnatenko25}. Furthermore, the
majority of studies based on Krylov state complexity concentrate on
time-independent Hamiltonian evolutions, with the extension to a completely
nonstationary framework still in its nascent phase \cite{nizami23,takahashi25}%
. For these reasons, the findings we present here are even more significant.

\medskip

Despite these considerable advancements, several limitations and potential
improvements deserve consideration. From the perspective of curvature
analysis, one might focus on examining the temporal profiles of magnetic
fields that could more closely resemble actual physical realizations within a
quantum laboratory. Nevertheless, one of the significant challenges in this
regard is the difficulty in acquiring precise analytical solutions to the
time-dependent Schr\"{o}dinger equation
\cite{landau32,zener32,rabi37,barnes12,barnes13,messina14,grimaudo18,elena20,grimaudo23}%
. Furthermore, although the formula for $\kappa_{\mathrm{AC}}^{2}$ is
theoretically applicable to any $d$-level quantum system evolving under a
nonstationary Hamiltonian, our investigation was limited to systems comprising
only two levels. In this simpler context, the comprehension and visual
depiction of Bloch vectors and Bloch spheres are relatively straightforward,
in contrast to more intricate, higher-dimensional scenarios. As we transition
from systems with merely two levels to those with additional dimensions, the
clarity of these visual representations diminishes. Indeed, quantum theory
reveals distinctive features in more complex systems, including the simplest
yet non-trivial case involving three levels, referred to as qutrits
\cite{kurzy11}. These unusual quantum phenomena complicate the understanding
of the geometric structure of quantum systems in higher dimensions
\cite{xie20,siewert21}. We suggest consulting Ref. \cite{gamel16} for a
thorough overview of how Bloch vectors are utilized to represent single-qubit,
single-qutrit, and two-qubit systems, employing matrices such as Pauli,
Gell-Mann, and Dirac matrices, respectively. From a complexity analysis
perspective, although we have proposed our own method for quantifying the
complexity of quantum evolutions, it is crucial to recognize the established
methodologies available in the literature, as noted in the Introduction. A
significant approach is the one based on Krylov complexity
\cite{chapman18,vijay20,ali20,iaconis21,vijay22,caputa22,craps24,rolph24,nath25,pg25}%
. This complexity measure serves as a vital metric for assessing the rate at
which quantum states propagate over time within the Krylov space derived from
the initial state or, alternatively, the speed at which operators disseminate
across the entire spectrum of potential operators during dynamic evolution. It
is anticipated that the late-time plateau of this metric will aid in
differentiating between integrable and chaotic dynamics; however, its
effectiveness in this regard is significantly influenced by the selection of
the initial seed \cite{pg25,craps25}. It would be valuable to conduct a
thorough comparative analysis between our complexity measure and the Krylov
state complexity in both stationary and nonstationary qubit dynamics contexts.
Nevertheless, this matter falls outside the current scope, and for now, we
shall postpone this intriguing inquiry to future scientific pursuits.

\medskip

To sum up, even though it has its present constraints, we truly believe that
our research will encourage other researchers and set the stage for more
in-depth studies on how geometry and quantum mechanics interact.

\begin{acknowledgments}
C.C. is grateful to the Griffiss Institute (Rome-NY) and to the United States
Air Force Research Laboratory (AFRL) Visiting Faculty Research Program (VFRP)
for providing support for this work. J.S. acknowledges support from the AFRL.
The authors appreciate the insightful conversations conducted with P. M.
Alsing. Any opinions, findings and conclusions or recommendations expressed in
this material are those of the authors and do not necessarily reflect the
views of the AFRL.
\end{acknowledgments}

\pagebreak

\appendix

\section{Complexity calculations}

In this appendix, we provide some technical details on the derivations of Eqs.
(\ref{com4}) and (\ref{com5}).

\subsection{Exponential growth}

In this first part of the appendix, we calculate the complexity of the quantum
evolution in the exponential growth scenario. Specifically, we consider the
case in which $\alpha\left(  t\right)  \overset{\text{def}}{=}\omega_{0}t$,
$\beta\left(  t\right)  \overset{\text{def}}{=}e^{\nu_{0}t}$, and $0\leq
t\leq\pi/(2\omega_{0})$. The state of the system at time $t$ can be recast as,%
\begin{align}
\left\vert \psi\left(  t\right)  \right\rangle  &  =\left\vert \psi\left(
\theta\left(  t\right)  \text{, }\varphi\left(  t\right)  \right)
\right\rangle \nonumber\\
&  =\cos\left[  \frac{\theta\left(  t\right)  }{2}\right]  \left\vert
0\right\rangle +e^{i\varphi\left(  t\right)  }\sin\left[  \frac{\theta\left(
t\right)  }{2}\right]  \left\vert 1\right\rangle \nonumber\\
&  =\cos\left[  \alpha\left(  t\right)  \right]  \left\vert 0\right\rangle
+e^{i\beta\left(  t\right)  }\sin\left[  \alpha\left(  t\right)  \right]
\left\vert 1\right\rangle \text{.} \label{arrive1}%
\end{align}
From Eq. (\ref{arrive1}), we arrive at $\theta\left(  t\right)  =2\alpha
\left(  t\right)  =2\omega_{0}t$ and $\varphi\left(  t\right)  =\beta\left(
t\right)  =e^{\nu_{0}t}$. Furthermore, since $0\leq t\leq\pi/(2\omega_{0})$,
we also have $0\leq\theta\leq\pi$ and $1\leq\varphi\leq e^{\frac{\pi}{2}%
\frac{\nu_{0}}{\omega_{0}}}$. Given this temporal interval and these bounds on
the parameters $\theta$ and $\varphi$, we can now proceed with the calculation
of the instantaneous, accessed, and accessible volumes $V(t)$, $\overline
{\mathrm{V}}$, and $\mathrm{V}_{\max}$, respectively. Recalling that volumes
are positively defined, the instantaneous volume $V(t)$ becomes
\begin{equation}
V(t)=\left\vert \int_{\varphi\left(  0\right)  }^{\varphi\left(  t\right)
}\int_{\theta\left(  0\right)  }^{\theta\left(  t\right)  }\sqrt
{g_{\mathrm{FS}}\left(  \theta\text{, }\varphi\right)  }d\theta d\varphi
\right\vert =\frac{\left[  1-\cos\left(  2\omega_{0}t\right)  \right]  \left(
e^{\nu_{0}t}-1\right)  }{4}\text{.} \label{ivolume}%
\end{equation}
From Eq. (\ref{ivolume}), the accessed volume $\overline{\mathrm{V}}$ reduces
to%
\begin{align}
\overline{\mathrm{V}}  &  =\frac{2\omega_{0}}{\pi}\int_{0}^{\frac{\pi}%
{2\omega_{0}}}V(t)dt\nonumber\\
&  =\frac{2\omega_{0}}{\pi}\frac{1}{4}\int_{0}^{\frac{\pi}{2\omega_{0}}%
}\left[  1-\cos\left(  2\omega_{0}t\right)  \right]  \left(  e^{\nu_{0}%
t}-1\right)  dt\nonumber\\
&  =\frac{1}{4\pi}\int_{0}^{\pi}\left[  1-\cos(x)\right]  \left(  e^{\frac
{1}{2}\frac{\nu_{0}}{\omega_{0}}x}-1\right)  dx\nonumber\\
&  =\frac{1}{4\pi}\int_{0}^{\pi}\left[  1-\cos(x)\right]  e^{\frac{1}{2}%
\frac{\nu_{0}}{\omega_{0}}x}dx-\frac{1}{4\pi}\int_{0}^{\pi}\left[
1-\cos(x)\right]  dx\nonumber\\
&  =\frac{1}{4\pi}\int_{0}^{\pi}\left[  1-\cos(x)\right]  e^{\frac{1}{2}%
\frac{\nu_{0}}{\omega_{0}}x}dx-\frac{1}{4}\text{,} \label{iv2}%
\end{align}
where $x\overset{\text{def}}{=}2\omega_{0}t$. Finally, calculating the
integral on the right-hand-side of Eq. (\ref{iv2}), we obtain%
\begin{equation}
\overline{\mathrm{V}}=\overline{\mathrm{V}}\left(  \frac{\nu_{0}}{\omega_{0}%
}\right)  \overset{\text{def}}{=}\frac{\left[  \left(  \frac{\nu_{0}}%
{\omega_{0}}\right)  ^{2}+2\right]  e^{\frac{\pi}{2}\frac{\nu_{0}}{\omega_{0}%
}}-2}{\pi\left(  \frac{\nu_{0}}{\omega_{0}}\right)  \left[  \left(  \frac
{\nu_{0}}{\omega_{0}}\right)  ^{2}+4\right]  }-\frac{1}{4}\text{.} \label{iv3}%
\end{equation}
Lastly, the accessible volume $\mathrm{V}_{\max}$ reduces to%
\begin{align}
\mathrm{V}_{\max}  &  =\left\vert \int_{\varphi_{\min}}^{\varphi_{\max}}%
\int_{\theta_{\min}}^{\theta_{\max}}\sqrt{g_{\mathrm{FS}}\left(  \theta\text{,
}\varphi\right)  }d\theta d\varphi\right\vert \nonumber\\
&  =\left\vert \int_{1}^{e^{\frac{\pi}{2}\frac{\nu_{0}}{\omega_{0}}}}\int%
_{0}^{\pi}\sqrt{g_{\mathrm{FS}}\left(  \theta\text{, }\varphi\right)  }d\theta
d\varphi\right\vert \nonumber\\
&  =\frac{1}{4}\left\vert \left(  \int_{0}^{\pi}\sin(\theta)d\theta\right)
\left(  \int_{1}^{e^{\frac{\pi}{2}\frac{\nu_{0}}{\omega_{0}}}}d\varphi\right)
\right\vert \nonumber\\
&  =\frac{e^{\frac{\pi}{2}\frac{\nu_{0}}{\omega_{0}}}-1}{2}\text{,}%
\end{align}
that is,%
\begin{equation}
\mathrm{V}_{\max}=\mathrm{V}_{\max}\left(  \frac{\nu_{0}}{\omega_{0}}\right)
\overset{\text{def}}{=}\frac{e^{\frac{\pi}{2}\frac{\nu_{0}}{\omega_{0}}}-1}%
{2}\text{.} \label{iv4}%
\end{equation}
Finally, combining Eqs. (\ref{iv3}) and (\ref{iv4}), we arrive at the
complexity of this particular quantum evolution. We get,%
\begin{equation}
\mathrm{C}=\frac{\mathrm{V}_{\max}-\mathrm{V}}{\mathrm{V}_{\max}}=\frac
{\frac{e^{\frac{\pi}{2}\frac{\nu_{0}}{\omega_{0}}}-1}{2}-\left(  \frac{\left[
\left(  \frac{\nu_{0}}{\omega_{0}}\right)  ^{2}+2\right]  e^{\frac{\pi}%
{2}\frac{\nu_{0}}{\omega_{0}}}-2}{\pi\left(  \frac{\nu_{0}}{\omega_{0}%
}\right)  \left[  \left(  \frac{\nu_{0}}{\omega_{0}}\right)  ^{2}+4\right]
}-\frac{1}{4}\right)  }{\frac{e^{\frac{\pi}{2}\frac{\nu_{0}}{\omega_{0}}}%
-1}{2}}\text{,}%
\end{equation}
that is, after some algebraic manipulations,%
\begin{equation}
\mathrm{C}=\mathrm{C}\left(  \frac{\nu_{0}}{\omega_{0}}\right)
\overset{\text{def}}{=}\frac{\left[  2\pi\left(  \frac{\nu_{0}}{\omega_{0}%
}\right)  ^{3}-4\left(  \frac{\nu_{0}}{\omega_{0}}\right)  ^{2}+8\pi\frac
{\nu_{0}}{\omega_{0}}-8\right]  e^{\frac{\pi}{2}\frac{\nu_{0}}{\omega_{0}}%
}-\left[  \pi\left(  \frac{\nu_{0}}{\omega_{0}}\right)  ^{3}+4\pi\frac{\nu
_{0}}{\omega_{0}}-8\right]  }{\left[  2\pi\left(  \frac{\nu_{0}}{\omega_{0}%
}\right)  ^{3}+8\pi\frac{\nu_{0}}{\omega_{0}}\right]  e^{\frac{\pi}{2}%
\frac{\nu_{0}}{\omega_{0}}}-\left[  2\pi\left(  \frac{\nu_{0}}{\omega_{0}%
}\right)  ^{3}+8\pi\frac{\nu_{0}}{\omega_{0}}\right]  }\text{.}
\label{yoyoone}%
\end{equation}
Note that setting $\xi\overset{\text{def}}{=}\nu_{0}/\omega_{0}$, we have%
\begin{equation}
\lim_{\xi\rightarrow\infty}\mathrm{C}\left(  \xi\right)  =\lim_{\xi
\rightarrow\infty}\frac{\left(  2\pi\xi^{3}-4\xi^{2}+8\pi\xi-8\right)
e^{\frac{\pi}{2}\xi}-\left(  \pi\xi^{3}+4\pi\xi-8\right)  }{\left(  2\pi
\xi^{3}+8\pi\xi\right)  e^{\frac{\pi}{2}\xi}-\left(  2\pi\xi^{3}+8\pi
\xi\right)  }=1\text{,} \label{w1}%
\end{equation}
and,%
\begin{equation}
\lim_{\xi\rightarrow0}\mathrm{C}\left(  \xi\right)  =\lim_{\xi\rightarrow
0}\frac{\left(  2\pi\xi^{3}-4\xi^{2}+8\pi\xi-8\right)  e^{\frac{\pi}{2}\xi
}-\left(  \pi\xi^{3}+4\pi\xi-8\right)  }{\left(  2\pi\xi^{3}+8\pi\xi\right)
e^{\frac{\pi}{2}\xi}-\left(  2\pi\xi^{3}+8\pi\xi\right)  }=\frac{3\pi^{2}%
-4}{4\pi^{2}}\sim\allowbreak0.65\text{.} \label{w2}%
\end{equation}
Therefore, in the limit of $\nu_{0}/\omega_{0}\rightarrow0$ of Eq. (\ref{w2}),
the limiting value of the complexity in the exponential growth case reduces to
the one we obtained in the linear growth case. Indeed, keeping in mind the
specifics of the linear growth case, this behavior can be explained by
observing that when $\nu_{0}/\omega_{0}\rightarrow0$, the exponential growth
case is characterized by $\varphi\left(  t\right)  \approx1+\nu_{0}t$ with
$1\leq\varphi\left(  t\right)  \lesssim1+(\pi/2)(\nu_{0}/\omega_{0})$.
Moreover, when $\nu_{0}/\omega_{0}\rightarrow\infty$ as in Eq. (\ref{w1}),
$\overline{\mathrm{V}}\approx\left(  \nu_{0}/\omega_{0}\right)  ^{-1}%
e^{\frac{\pi}{2}\frac{\nu_{0}}{\omega_{0}}}$, and $\mathrm{V}_{\max}\approx
e^{\frac{\pi}{2}\frac{\nu_{0}}{\omega_{0}}}$. Therefore, $\overline
{\mathrm{V}}/\mathrm{V}_{\max}\approx\left(  \nu_{0}/\omega_{0}\right)
^{-1}\rightarrow0$ as $\nu_{0}/\omega_{0}$ diverges. This type of behavior
specifies a maximal complexity scenario in which \textrm{C} approaches its
maximal value one.

\subsection{Exponential decay}

$\allowbreak$In this second part of the appendix, we evaluate the complexity
of the quantum evolution in the exponential decay case. Specifically, we
assume to consider the case where $\alpha\left(  t\right)  \overset{\text{def}%
}{=}\omega_{0}t$, $\beta\left(  t\right)  \overset{\text{def}}{=}e^{-\nu_{0}%
t}$, and $0\leq t\leq\pi/(2\omega_{0})$. The state of the system at time $t$
can be written as,%
\begin{align}
\left\vert \psi\left(  t\right)  \right\rangle  &  =\left\vert \psi\left(
\theta\left(  t\right)  \text{, }\varphi\left(  t\right)  \right)
\right\rangle \nonumber\\
&  =\cos\left[  \frac{\theta\left(  t\right)  }{2}\right]  \left\vert
0\right\rangle +e^{i\varphi\left(  t\right)  }\sin\left[  \frac{\theta\left(
t\right)  }{2}\right]  \left\vert 1\right\rangle \nonumber\\
&  =\cos\left[  \alpha\left(  t\right)  \right]  \left\vert 0\right\rangle
+e^{i\beta\left(  t\right)  }\sin\left[  \alpha\left(  t\right)  \right]
\left\vert 1\right\rangle \text{.} \label{arrive1B}%
\end{align}
From Eq. (\ref{arrive1B}), we get $\theta\left(  t\right)  =2\alpha\left(
t\right)  =2\omega_{0}t$ and $\varphi\left(  t\right)  =\beta\left(  t\right)
=e^{-\nu_{0}t}$. Furthermore, since $0\leq t\leq\pi/(2\omega_{0})$, we also
have $0\leq\theta\leq\pi$ and $e^{-\frac{\pi}{2}\frac{\nu_{0}}{\omega_{0}}%
}\leq\varphi\leq1$. Given this temporal interval and these bounds on the
parameters $\theta$ and $\varphi$, we can now proceed with the computation of
the instantaneous, accessed, and accessible volumes $V(t)$, $\overline
{\mathrm{V}}$, and $\mathrm{V}_{\max}$, respectively. Recalling that volumes
are positive quantities, the instantaneous volume $V(t)$ becomes
\begin{equation}
V(t)=\left\vert \int_{\varphi\left(  0\right)  }^{\varphi\left(  t\right)
}\int_{\theta\left(  0\right)  }^{\theta\left(  t\right)  }\sqrt
{g_{\mathrm{FS}}\left(  \theta\text{, }\varphi\right)  }d\theta d\varphi
\right\vert =\frac{\left[  1-\cos\left(  2\omega_{0}t\right)  \right]  \left(
1-e^{-\nu_{0}t}\right)  }{4}\text{.} \label{ivolumeB}%
\end{equation}
From Eq. (\ref{ivolumeB}), the accessed volume $\overline{\mathrm{V}}$ is
given by%
\begin{align}
\overline{\mathrm{V}}  &  =\frac{2\omega_{0}}{\pi}\int_{0}^{\frac{\pi}%
{2\omega_{0}}}V(t)dt\nonumber\\
&  =\frac{2\omega_{0}}{\pi}\frac{1}{4}\int_{0}^{\frac{\pi}{2\omega_{0}}%
}\left[  1-\cos\left(  2\omega_{0}t\right)  \right]  \left(  1-e^{-\nu_{0}%
t}\right)  dt\nonumber\\
&  =\frac{1}{4\pi}\int_{0}^{\pi}\left[  1-\cos(x)\right]  \left(
1-e^{-\frac{1}{2}\frac{\nu_{0}}{\omega_{0}}x}\right)  dx\nonumber\\
&  =\frac{1}{4\pi}\int_{0}^{\pi}\left[  1-\cos(x)\right]  dx-\frac{1}{4\pi
}\int_{0}^{\pi}\left[  1-\cos(x)\right]  e^{-\frac{1}{2}\frac{\nu_{0}}%
{\omega_{0}}x}dx\nonumber\\
&  =\frac{1}{4}-\frac{1}{4\pi}\int_{0}^{\pi}\left[  1-\cos(x)\right]
e^{-\frac{1}{2}\frac{\nu_{0}}{\omega_{0}}x}dx\text{,} \label{iv2B}%
\end{align}
where $x\overset{\text{def}}{=}2\omega_{0}t$. Finally, evaluating the integral
on the right-hand-side of Eq. (\ref{iv2B}), we get%
\begin{equation}
\overline{\mathrm{V}}=\overline{\mathrm{V}}\left(  \frac{\nu_{0}}{\omega_{0}%
}\right)  \overset{\text{def}}{=}\frac{1}{4}-\frac{2-\left[  \left(  \frac
{\nu_{0}}{\omega_{0}}\right)  ^{2}+2\right]  e^{-\frac{\pi}{2}\frac{\nu_{0}%
}{\omega_{0}}}}{\pi\left(  \frac{\nu_{0}}{\omega_{0}}\right)  \left[  \left(
\frac{\nu_{0}}{\omega_{0}}\right)  ^{2}+4\right]  }\text{.} \label{iv3B}%
\end{equation}
Ultimately, the accessible volume $\mathrm{V}_{\max}$ becomes%
\begin{align}
\mathrm{V}_{\max}  &  =\left\vert \int_{\varphi_{\min}}^{\varphi_{\max}}%
\int_{\theta_{\min}}^{\theta_{\max}}\sqrt{g_{\mathrm{FS}}\left(  \theta\text{,
}\varphi\right)  }d\theta d\varphi\right\vert \nonumber\\
&  =\left\vert \int_{e^{-\frac{\pi}{2}\frac{\nu_{0}}{\omega_{0}}}}^{1}\int%
_{0}^{\pi}\sqrt{g_{\mathrm{FS}}\left(  \theta\text{, }\varphi\right)  }d\theta
d\varphi\right\vert \nonumber\\
&  =\frac{1}{4}\left\vert \left(  \int_{0}^{\pi}\sin(\theta)d\theta\right)
\left(  \int_{e^{-\frac{\pi}{2}\frac{\nu_{0}}{\omega_{0}}}}^{1}d\varphi
\right)  \right\vert \nonumber\\
&  =\frac{1-e^{-\frac{\pi}{2}\frac{\nu_{0}}{\omega_{0}}}}{2}\text{,}%
\end{align}
that is,%
\begin{equation}
\mathrm{V}_{\max}=\mathrm{V}_{\max}\left(  \frac{\nu_{0}}{\omega_{0}}\right)
\overset{\text{def}}{=}\frac{1-e^{-\frac{\pi}{2}\frac{\nu_{0}}{\omega_{0}}}%
}{2}\text{.} \label{iv4B}%
\end{equation}
Finally, making use of Eqs. (\ref{iv3B}) and (\ref{iv4B}), we obtain the
complexity of this particular quantum evolution. We arrive at%
\begin{equation}
\mathrm{C}=\frac{\mathrm{V}_{\max}-\mathrm{V}}{\mathrm{V}_{\max}}=\frac
{\frac{1-e^{-\frac{\pi}{2}\frac{\nu_{0}}{\omega_{0}}}}{2}-\left(  \frac{1}%
{4}-\frac{2-\left[  \left(  \frac{\nu_{0}}{\omega_{0}}\right)  ^{2}+2\right]
e^{-\frac{\pi}{2}\frac{\nu_{0}}{\omega_{0}}}}{\pi\left(  \frac{\nu_{0}}%
{\omega_{0}}\right)  \left[  \left(  \frac{\nu_{0}}{\omega_{0}}\right)
^{2}+4\right]  }\right)  }{\frac{1-e^{-\frac{\pi}{2}\frac{\nu_{0}}{\omega_{0}%
}}}{2}}\text{,}%
\end{equation}
that is, after some algebraic manipulations,%
\begin{equation}
\mathrm{C}=\mathrm{C}\left(  \frac{\nu_{0}}{\omega_{0}}\right)
\overset{\text{def}}{=}\frac{\left[  \pi\left(  \frac{\nu_{0}}{\omega_{0}%
}\right)  ^{3}+4\pi\left(  \frac{\nu_{0}}{\omega_{0}}\right)  +8\right]
e^{\frac{\pi}{2}\frac{\nu_{0}}{\omega_{0}}}-\left[  2\pi\left(  \frac{\nu_{0}%
}{\omega_{0}}\right)  ^{3}+4\left(  \frac{\nu_{0}}{\omega_{0}}\right)
^{2}+8\pi\left(  \frac{\nu_{0}}{\omega_{0}}\right)  +8\right]  }{\left[
2\pi\left(  \frac{\nu_{0}}{\omega_{0}}\right)  ^{3}+8\pi\left(  \frac{\nu_{0}%
}{\omega_{0}}\right)  \right]  e^{\frac{\pi}{2}\frac{\nu_{0}}{\omega_{0}}%
}-\left[  2\pi\left(  \frac{\nu_{0}}{\omega_{0}}\right)  ^{3}+8\pi\left(
\frac{\nu_{0}}{\omega_{0}}\right)  \right]  }\text{.} \label{yoyotwo}%
\end{equation}
Observe that putting $\xi\overset{\text{def}}{=}\nu_{0}/\omega_{0}$, we obtain%
\begin{equation}
\lim_{\xi\rightarrow\infty}\mathrm{C}\left(  \xi\right)  =\lim_{\xi
\rightarrow\infty}\frac{\left(  \pi\xi^{3}+4\pi\xi+8\right)  e^{\frac{\pi}%
{2}\xi}-\left(  2\pi\xi^{3}+4\xi^{2}+8\pi\xi+8\right)  }{\left(  2\pi\xi
^{3}+8\pi\xi\right)  e^{\frac{\pi}{2}\xi}-\left(  2\pi\xi^{3}+8\pi\xi\right)
}=\frac{1}{2}\text{,} \label{u1}%
\end{equation}
and,%
\begin{equation}
\lim_{\xi\rightarrow0}\mathrm{C}\left(  \xi\right)  =\lim_{\xi\rightarrow
0}\frac{\left(  2\pi\xi^{3}-4\xi^{2}+8\pi\xi-8\right)  e^{\frac{\pi}{2}\xi
}-\left(  \pi\xi^{3}+4\pi\xi-8\right)  }{\left(  2\pi\xi^{3}+8\pi\xi\right)
e^{\frac{\pi}{2}\xi}-\left(  2\pi\xi^{3}+8\pi\xi\right)  }=\frac{3\pi^{2}%
-4}{4\pi^{2}}\sim0.65\text{.} \label{u2}%
\end{equation}
Therefore, in the limit of $\nu_{0}/\omega_{0}\rightarrow0$ of Eq. (\ref{u2}),
the limiting value of the complexity in the exponential decay case equals the
one we obtained in the linear growth case. Indeed, taking into consideration
the details of the linear growth case, this behavior can be grasped by noting
that when $\nu_{0}/\omega_{0}\rightarrow0$, the exponential decay case is
specified by $\varphi\left(  t\right)  \approx1-\nu_{0}t$ with $1-(\pi
/2)(\nu_{0}/\omega_{0})\leq\varphi\left(  t\right)  \lesssim1$. Moreover, when
$\nu_{0}/\omega_{0}\rightarrow\infty$ as in Eq. (\ref{u1}), $\varphi\left(
t\right)  $ becomes constant and we recover the same complexity value
calculated in the no growth case.

With these observations and with the derivations of Eqs. (\ref{yoyoone}) and
(\ref{yoyotwo}), we end this appendix.

\bigskip

\end{document}